\documentclass[a4paper,12pt]{amsart}
\usepackage[utf8]{inputenc}
\usepackage[all]{xy}
\usepackage{amssymb}
\usepackage{csquotes}
\usepackage{graphicx}
\usepackage{enumerate}
\usepackage{amsaddr}
\usepackage[nocompress]{cite}
\newtheorem{lemma}{Lemma}
\newtheorem{remark}{Remark}
\newtheorem{theorem}{Theorem}
\newtheorem{proposition}{Proposition}
\newtheorem{definition}{Definition}
\DeclareMathOperator{\Tr}{Tr}

\title{Transforming cumulative hazard estimates}
\author{Pål C. Ryalen, Mats J. Stensrud, and Kjetil Røysland}
\address{Department of Biostatistics, University of Oslo, Domus Medica Gaustad, Sognsvannsveien 9, 0372 Oslo, Norway}
\date{\today}
\MakeOuterQuote{"}\EnableQuotes
\DeclareUnicodeCharacter{00A0}{ }

\begin{document}

\begin{abstract}
Time to event outcomes are often evaluated on the hazard scale, but interpreting hazards may be difficult. Recently, there has been concern in the causal inference literature that hazards actually have a built in selection-effect that prevents simple causal interpretations. This is even a problem in randomized controlled trials, where hazard ratios have become a standard measure of treatment effects. Modeling on the hazard scale is nevertheless convenient, e.g. to adjust for covariates. Using hazards for intermediate calculations may therefore be desirable. Here, we provide a generic method for transforming hazard estimates consistently to other scales at which these built in selection effects are avoided. The method is based on differential equations, and generalize a well known relation between the Nelson-Aalen and Kaplan-Meier estimators. Using the martingale central limit theorem we also find that covariances can be estimated consistently for a large class of estimators, thus allowing for rapid calculations of confidence intervals. Hence, given cumulative hazard estimates based on e.g. Aalen's additive hazard model, we can obtain many other parameters without much more effort. We present several examples and associated estimators. Coverage and convergence speed are explored using simulations, suggesting that reliable estimates can be obtained in real-life scenarios.
\end{abstract}
\maketitle

%\blfootnote{Corresponding author: Pål C. Ryalen; E-mail: p.c.ryalen@medisin.uio.no}
\section{Outline}
Applied researchers are often faced with time to event outcomes. Causal inference for such outcomes is challenging due to the dynamic aspect of time, with possibly time-varying effects and biases. In particular, the proportional hazard model is the standard approach for analyzing survival data in medicine, but assuming proportional hazards cannot be justified in many real-life scenarios. Furthermore, since estimating hazards involves conditioning on recent survival, the presence of a seemingly innocent unobserved heterogeneity means that we will condition on a so-called collider, and will therefore activate non-causal pathways from the exposure to the risk of an event at short term \cite{aalen2015does}. These commonly used effect estimates can therefore often not be interpreted causally as short-term risks \cite{robins1989probability, greenland1996absence,stensrud2017exploring,aalen2015does,aalen2008survival,hernan2010hazards}.

However, to evaluate time to event outcomes, we do not need to assess hazards per se. We may rather be interested in effect measures on e.g. the survival scale. Such measures could be easier to interpret, and allow for causal evaluations \cite{hernan2010hazards,hernan2017causal}. Still, modeling the hazard scale may be useful as an intermediate step, e.g. to adjust for covariates.

Additive hazard models seem to become more popular, at least in the causal inference literature \cite{martinussen2011estimation, lange2011direct}. In contrast to the proportional hazards model, the additive models are collapsible and they easily allow for time-dependent effects. The additive models are also less prone to selection biases when unmeasured factors follow linear structural models \cite{strohmaier2015dynamic, martinussen2013collapsibility}.
On the other hand, additive models have been criticized because the effect estimates may be harder to understand: Often effect estimates are directly plotted as cumulative hazard curves, which seems to be unsatisfactory to applied researchers \cite{bradburn2003survival}. Indeed, the cumulative hazard curves may neither have an immediate causal interpretation.

With this in mind, we aim to transform cumulative hazard estimates to other scales by exploiting structure imposed by ordinary differential equations. 
We will estimate parameters that solve such equations, which are typically driven by cumulative hazards. 
The estimators we consider are solutions of naturally associated stochastic differential equations that are straight forward to solve numerically on a computer. 
We are mostly concerned with such equations that are driven by cumulative hazards, but our method is not at all limited to this setting. Our approach enables us to apply a simple variant of the delta-method to prove that applying our transformations
to consistent cumulative hazard estimators also yield consistent estimators. Furthermore, we are  able to provide estimates of the asymptotic variance. In fact, this is much simpler and allows us to go further than  the functional delta-method suggested by Richard D. Gill et al. \cite{ gill1989non, Andersen}, since the latter involves topologies that sometimes need to be carefully designed for each example, 
which may be hard to handle in practice; see e.g. \cite{van1998asymptotic}.  

\section{Parameters}
\label{section:parameters}

The parameters we will consider are functions on a finite interval $[0,\mathcal T]$ that are solutions to an ordinary differential equation system on the form 
\begin{equation} \label{eq:ODEdef}
        \pmb{X}_t = \pmb{X}_0 + \int_0^t F(\pmb{X}_{s-}) d\pmb{A}_{s}.
\end{equation}
Here $\pmb{A}$ is a continuous $k$-dimensional function of bounded variation, and $F : \mathbb R^n \rightarrow
M_{n,k}( \mathbb R)$ is sufficiently smooth and satisfies a linear growth bound.
There are several examples of parameters on this form in the survival analysis literature, but \eqref{eq:ODEdef} has not received much attention. In the following we show examples of parameters that can be written in this way.

\subsection{Survival curves}
  Let $T \geq 0$ be a survival time and let 
  $S_t := P( T \geq t)$, i.e. $S$ is the survival function. 
  If $A_t$ denotes the
  corresponding cumulative hazard, then we have the  relation 
  \begin{align*}
          S_t = S_0  - \int_0^t S_{s-} d A_s. 
  \end{align*}

\subsection{Relative survival}
When assessing effects of exposures, it is intriguing to consider the probability of survival directly, rather than estimating e.g.\ hazard ratios. The relative survival is routinely used in clinical medicine \cite{perme2012estimation}. In particular, five year relative survival rates are standard measures of the survival in cancer patients vs. the general population, and such numbers are conventionally reported from cancer registries. Suppose that we are given two groups $1$ and $0$, and we want to identify 
the relative survival $RS_t := S^1_t /S_t^0$ of the positive group, relatively to the
negative group at time $t$. If we let $A^1$ and $A^0$ denote the corresponding 
cumulative hazards, then 
a simple calculation of derivatives  shows that 
\begin{align*}
         RS_t 
        = 
        RS_0 
+ \int_0^t   
\begin{pmatrix} 
-RS_{s-}   & RS_{s-}   \end{pmatrix}
d\begin{pmatrix} A_s^1 \\ A_s^0 \end{pmatrix}.
\end{align*}
%For the general theory in this paper to hold, $S^0_{s-}$ must be such that the relative survival is bounded.

\subsection{Mean and restricted mean survival}
Instead of studying the relative survival as a function of time, we may be interested in the mean survival in different populations. Indeed, the mean survival is simply the area under the survival curve. In practice, however, the tail of the estimated survival function will strongly influence the integral of the estimated survival curve, and the mean survival estimates may be unreliable. As a remedy, it is possible to study the restricted mean survival $E\big[ T  I(T \leq \tau)\big]$, that is the area under the survival curve up to some (restricted) time $\tau$, $R_{\tau}$. It may be interpreted as the life expectancy between $t=0$ and $\tau$. Advocates claim that the restricted mean survival to a prespecified $\tau$ should be reported in clinical trials \cite{royston2011use,royston2013restricted,trinquart2016comparison}, in particular when the proportional hazards assumptions is invalid. Indeed, $R_{\tau}$ is readily found by solving the system
\begin{align*}
    \begin{pmatrix}
    R_{\tau} \\
    S_{\tau}
    \end{pmatrix} &=
    \begin{pmatrix}
    0 \\
    1
    \end{pmatrix}+
    \int_0^{\tau}
    \begin{pmatrix}
    S_{s} & 0 \\
    0 & -S_{s-}
    \end{pmatrix}
    d\begin{pmatrix}
    s \\
    A_s
    \end{pmatrix},
\end{align*}
where $A$ is the cumulative hazard for death.

\subsection{Life expectancy difference and life expectancy ratio}
The life expectancy difference (LED) and life expectancy ratio (LER) can be useful to compare (restricted) mean survivals in different groups\cite{dehbi2017life}. The LED between two groups 1 and 2 is the difference between their restricted mean survival times. It solves the system

\begin{align*}
    \begin{pmatrix}
    \text{LED}_{t}^{12} \\
    S_{t}^1 \\
    S_{t}^2
    \end{pmatrix} &=
    \begin{pmatrix}
    0 \\
    1 \\
    1
    \end{pmatrix}+
    \int_0^{t}
    \begin{pmatrix}
    S_{s}^1 - S_{s}^2 & 0 & 0  \\
    0 & -S_{s-}^1 & 0  \\
    0 & 0 & -S_{s-}^2 
    \end{pmatrix}
    d\begin{pmatrix}
    s \\
    A_s^1\\
    A_s^2
    \end{pmatrix}.
\end{align*}

Here, $S^1,S^2$ and $A^1,A^2$ are the survival, and cumulative hazards in group $1$ and $2$ respectively. The life expectancy ratio is the ratio of the restricted mean survival times. It solves the system

\begin{align*}
    \begin{pmatrix}
    \text{LER}_{t}^{12} \\
    S_{t}^1 \\
    S_{t}^2 \\
    R_t^1 \\
    R_t^2
    \end{pmatrix} &=
    \begin{pmatrix}
    1 \\
    1 \\
    1 \\
    0 \\
    0
    \end{pmatrix}+
    \int_0^{t}
    \begin{pmatrix}
    \frac{S_{s-}^1R_{s-}^2 - S_{s-}^2R_{s-}^1}{(R_{s-}^2)^2} & 0 & 0  \\
    0 & -S_{s-}^1 & 0  \\
    0 & 0 & -S_{s-}^2  \\
    S_{s-}^1 & 0 & 0 \\
    S_{s-}^2 & 0 & 0
    \end{pmatrix}
    d\begin{pmatrix}
    s \\
    A_s^1\\
    A_s^2
    \end{pmatrix},
\end{align*}
where $R_{s-}^2$ cannot be too small.

%When proportional hazards assumptions is invalid, the the area under the survival curve. When integrating the estimated survival curve, however, the tail of the estimated survival function will have a large influence.

%restricted mean survival be useful, in particular when the hazards are non proportional\cite{royston2011use, royston2013restricted. trinquart2016comparison}. 

%It is often of interest to study the mean survival, which is nothing but the area under the survival curve. When integrating the estimated survival curve, however, the tail of the estimated survival function will have a large influence. It could then be compelling to consider the restricted mean survival; the area under the survival curve up to some (restricted) time $\tau$, $R_{\tau}$.

\subsection{Cumulative incidence and competing risks}
In practice we are often interested in the time to a particular event, but our study may be disrupted by other, \textit{competing} events. In medicine, we may e.g.\ be interested in the time to the onset of a disease $D$, but subjects may die before they develop $D$. If there are competing risks, it is incorrect to assume a one-to-one relation between cause-specific hazards and the cumulative incidence\cite{andersen2012competing}; if we e.g.\ treat death as censoring, we obtain estimates from a hypothetical population in which subjects cannot die without $D$.

Alternatively, we may include the hazards of the competing events into our model. Consider a situation with $k$ competing causes of death, with  cause-specific
cumulative hazards $A^1, \dots, A^k$. 
Following Andersen et.al \cite{andersen2002competing}, we see that 
\begin{equation}
        \begin{pmatrix}
                S_t \\ 
                C^1_t \\ \vdots \\C^k_t
        \end{pmatrix}
        = \begin{pmatrix}
                1 \\0 \\\vdots\\ 0
        \end{pmatrix}
        +\int_0^t          
        \begin{pmatrix}
        -S_{s-} & -S_{s-}  & \dots & -S_{s-} \\
        S_{s-}  & 0 & \dots & 0\\
        0 & \dots & \ddots  & S_{s-}
        \end{pmatrix}
        d\begin{pmatrix}
         A^1_s \\ \vdots \\ A^k_s       
        \end{pmatrix}
\end{equation}
describes all the cause-specific cumulative incidences $C^1_t, \dots , C^k_t$,
subject to the remaining competing risks.

\subsection{Mean frequency function for recurrent events}

Consider individuals that may experience a number of recurrent events, e.g. hospital visits $H$, while being at risk of experiencing some terminating event, e.g. death $D$. We let $N_t^H$ be the counting process that counts the number of recurrent events up to time $t$, and $A^H,A^D$ be the cumulative hazards for the recurrent and terminating event, respectively. The mean frequency function $K_t = E[N_{t\wedge D}^H]$ then solves the system
\begin{align}
    \begin{pmatrix}
    K_t \\
    S_t
    \end{pmatrix}&= 
    \begin{pmatrix}
    0 \\
    1
    \end{pmatrix}+
    \int_0^t
    \begin{pmatrix}
        S_{s-} & 0  \\
        0 & -S_{s-}
    \end{pmatrix}
    d\begin{pmatrix}
        A_s^H \\
        A_s^D
    \end{pmatrix}.
    \label{eq:recurrentEventSystem}
\end{align}
$A^H$ and $A^D$ can be estimated using Nelson-Aalen estimators, and we can estimate $K$ by plugging them into \eqref{eq:recurrentEventSystem}. Asymptotic results for this estimator have previously been described \cite{ghoshlin2000}. The consistency result will be a consequence from a more general results in our paper.

\subsection{Cumulative sensitivity and specificity after screening}
The effectiveness of disease screening is frequently studied in medicine. Often the screening is performed at $t = 0$, but the disease may advance or develop at times $t>0$. Hence, subjects may be followed over time to assess if events occur, e.g.\ the onset of disease. In such scenarios, conventional methods to assess sensitivity and specificity are inadequate, because they do not account for differences in follow-up times. We may rather express sensitivity and specificity as functions of time, as suggested in \cite{nygaard2014comparative}.
Let $A^1$ be the hazard of the disease event for the group with positive
screening results, let $A^0$ be the hazard of the negative ones, let $\beta$
denote the prevalence of the disease and let $Z$ denote the screening outcome.
We let $U_t$ denote the cumulative positive predictive value $ P \big( N_t > 0
\big| Z = 1 \big)$, let 
$V_t$ denote the  cumulative negative predictive value $P \big( N_t  = 0 \big| Z = 0 \big) 
$, let  
$W_t$ denote the cumulative sensitivity $ P \big( Z = 1    \big| N_t  > 0  \big) $ and 
let $X_t$ denote the cumulative specificity $P \big( Z = 0    \big| N_t  =  0
        \big)$. 
Applying Bayes' rule gives the equation
\begin{equation}
        \begin{pmatrix}
                U_t \\V_t \\W_t \\X_t  
        \end{pmatrix}=
         \begin{pmatrix}
                U_0 \\V_0 \\W_0 \\X_0  
        \end{pmatrix} + \int_0^t    \begin{pmatrix}
               1 - U_{s-}  & 0  \\
               0  & - V_{s-} \\
               \frac{ W^2_{s-} ( 1- V_{s-})( 1-
                        U_{s-}) }{\gamma U_{s-}^2}  
               &
                - \frac{ W^2_{s-} V_{s-}U_{s-} } { \gamma U_{s-}^2 } 
                \\ 
                \frac{ \gamma X_{s-}^2 (1-
                U_{s-}) }{V_{s-} }  & -
                \frac{\gamma X_{s-}^2 (1-
                U_{s-}) }{V_{s-} }
        \end{pmatrix}
        d\begin{pmatrix}
        	A^1_s \\A^0_s       
        \end{pmatrix},
\end{equation}
where we have introduced the odds $\gamma = \frac{ \beta} {1- \beta}$, and where $V_{s-}$ and $U_{s-}$ cannot be too small.

\section{Plugin estimators}
\label{seq:SDEPluginEstimators}
Our suggested estimators will build on a cumulative hazard estimator ${\pmb{\hat A}}^{(n)}$ for $\pmb{A}$ that is given by counting process integrals, i.e. such that
\begin{equation} \label{eq:countingProcessEstimator}
        {\pmb{\hat A}}^{(n)}_t = \int_0^t \pmb{G}_{s-}^{(n)} d \pmb{N} ^{(n)}_s, 
\end{equation}
where $\pmb{G}^{(n)} = \begin{pmatrix}   
\pmb g^{(n)1} & \dots \pmb g^{(n)l}
\end{pmatrix}
$ is a predictable $k \times l$ dimensional matrix-valued process, and $\pmb{N}^{(n)}$ is an adapted $l$-dimensional counting process. It should be noted that the results here is applicable for more general situations than \eqref{eq:countingProcessEstimator}. 
Suppose, however, that this estimator is consistent in the sense that 
\begin{equation} \label{eq:ConsistentHazard}
        \lim _{n \rightarrow \infty} P \bigg( \sup_{s \leq \mathcal{T}} \big| \pmb{ \hat A}_s^{(n)}
-\pmb A_s \big| \geq
\epsilon \bigg) = 0,
\end{equation}
for every $\epsilon > 0$, 
or such that  
\begin{equation}\label{eq:Wn}
      \pmb {W}^{(n)} :=   \sqrt n \big({\pmb{ \hat A}}^{(n)}- \pmb A \big)
\end{equation}
converges in law to an independent increments mean zero Gaussian local martingale $\pmb W$ (relative to the Skorohod-metric).  The latter is typically a consequence of the martingale central limit theorem, as in \cite[Theorem II.5.1]{Andersen} or  \cite[Theorem VIII.3.22]{JacodShiryaev}. Two standard examples in event history analysis where this holds are the Nelson-Aalen cumulative hazard estimator\cite[Theorem IV.1.1 and Theorem IV.1.2]{Andersen} and Aalen's additive hazard regression \cite[Theorem VII.4.1]{Andersen}.

In addition to consistency, there is another property of $\pmb{\hat{A}}^{(n)}$
that will be crucial in this setting. This is \emph{predictably uniformly
        tightness}, abbreviated P-UT. The exact definition of P-UT is given in
\cite[VI.6a]{JacodShiryaev}. However, as we will not need the full generality,
we can use the following Lemma to  determine if processes are P-UT in 
most situations we  encounter.

\begin{lemma} \label{lemma:P-UTexplained}
              Let  $\{\pmb Z_t^{(n)}\}_n$ be a sequence of semi-martingales  on
              $[0,\mathcal T]$, and  let 
              $\{\pmb \rho^{(n)}\}_n$ be predictable processes such that  every 
              $$\pmb M^{(n)}_t := \pmb Z^{(n)}_t - \int_0 ^t \pmb \rho_s^{(n)} ds$$
                              defines a square integrable local martingale, and
                              suppose that 
              \begin{enumerate}[i)]
                                          \item \label{eq:tightRho}
                                                  \begin{equation*}
                                      \lim_{J \rightarrow \infty} \sup_n P \bigg( \sup_{s \leq \mathcal T}  |
                              \pmb \rho_s^{(n)} |_1 \geq J   \bigg) = 0
                                      \text{ and}  
                      \end{equation*}

                       \item \label{enum:PUTMartingale} \begin{equation} \label{eq:PUTMart}
                                      \lim_{J \rightarrow \infty} \sup_n P
                              \bigg(   \text{Tr} 
                              \langle \pmb M^{(n)} \rangle_{\mathcal{T}}  \geq J   \bigg) = 0. 
                      \end{equation}

              \end{enumerate}
              Then  $\{\pmb Z_t^{(n)}\}_n$ is P-UT. 
          
\end{lemma}

We are now ready to define the estimators for parameters that are defined through \eqref{eq:ODEdef}.

\begin{definition}\label{definition:SDEPluginEstimator}
Let $\pmb{\hat X}_0^{(n)}$ be an estimator for $\pmb X_0$. 
The plugin-estimator $\pmb{\hat X}^{(n)}$ for 
$\pmb{X}$ that satisfies \eqref{eq:ODEdef} is the solution of the following stochastic differential equation: 
\begin{equation} \label{eq:SDEpluginEst}
        \pmb{\hat X}_t^{(n)} = \pmb{\hat X}_0^{(n)} + \int_0^t  F( \pmb{\hat X}_{s-}^{(n)}) d\pmb{\hat A}^{(n)}_s. 
\end{equation}
\end{definition}

Plugin estimators are relatively easy to implement on a computer due to their recursive
form.  If $\tau_1, \tau_2,  \dots $ denote the jump times of  $\pmb {\hat A}^{(n)}$, then we have that that 
\begin{equation}\label{eq:SDEpluginNotation}
        \pmb {\hat X}_t^{(n)} = \pmb {\hat X}_{\tau_{k-1}}^{(n)} + F(   \pmb {\hat X}_{\tau_{k-1}}^{(n)} )  \Delta 
        \pmb {\hat A}^{(n)}_{\tau_{k}},
\end{equation}
whenever 
$\tau_{k} \leq t < \tau_{k+1}$. 
These estimators are also consistent in many situations. This is formulated as a Theorem, and is proved in the Appendix. 

\begin{theorem} \label{theorem:ODE}
       Suppose that 
       \begin{enumerate}[(i)]
       \item 
       $\pmb X$ satisfies the ordinary differential equation 
       \eqref{eq:ODEdef}, where 
       $F$ is locally Lipschitz-continuous, and satisfies the linear growth condition on a domain that contains $\{ \pmb X_t |t \in [0,\mathcal T]  \}$,
   \item \label{enumerate:SDEConvConsistency}
   $\{ {\pmb {\hat A}^{(n)}}\}_n$  and  
   $\{ {\pmb {\hat X}^{(n)}_0}\}_n$ are consistent, i.e. 
   \begin{align*}
        \lim_{n \rightarrow \infty} P \bigg( \sup_{s \leq \mathcal T} \big| 
\pmb {\hat A}^{(n)} _s - \pmb {A}_s  \big|
\geq \epsilon   \bigg)  = 0   \text{ and } 
                 \lim_{n \rightarrow \infty} P \bigg( \big| 
\pmb {\hat X}^{(n)}_0 -  \pmb X_0  \big|
\geq \epsilon   \bigg)  = 0, 
        \end{align*}
        for every $\epsilon >0$, 
   \item \label{enum:SDETightness} $\pmb {\hat A}^{(n)}$ is P-UT (see
           Proposition \ref{prop:AalenPUT} for additive hazards),
   %satisfies the conditions of Lemma \ref{lemma:P-UTexplained}, 
   
   %For each $t \in [0, \mathcal T]$,
%\begin{equation} \label{eq:totvarEst}
%        \{ \sum_{i =1 }^l \int_0 ^t \big|\pmb g_{s-}^{(n)i} \big|_1 dN_s^{(n)i}  \} _n    
%\end{equation}
%defines a tight family of finite random variables. 
       \end{enumerate}

       Then  
        \begin{equation}
                 \lim_{n \rightarrow \infty} P \bigg( \sup_{s \leq \mathcal T} \big| 
          {\pmb {\hat X}_s^{(n)}} - \pmb{ X}_s  \big|
\geq \epsilon   \bigg) = 0 
        \end{equation}
        for every $\epsilon > 0$, 
        i.e. 
        $\{ \pmb {\hat X}^{(n)}\}_n$ defines a consistent estimator of $\pmb
        X$. 
                      \end{theorem}

\begin{remark}\label{remark:deterministicCountingProcess}
    We can also handle the situation where $A_t = t$. %we have a sequence $\{\hat A^{(n)}\}_n$ that also satisfies \eqref{eq:ConsistentHazard}.  

    Consider a one-dimensional deterministic counting process  $N^{(n)}$ with jumping times $\tau_1^n < \tau_2^n <  \dots $
    such that 
    \begin{equation} \label{eq:grid}
       \lim_{n} \sup_{k} \big|  \tau_k^n - \tau_{k-1}^n \big|  = 0.
    \end{equation} Let
    \begin{equation*}
     h_t^{(n)} := \begin{cases}
     0, & t = 0\\
     \sup \{ \epsilon \geq 0  |N_{t - \epsilon }^{(n)}  = N_{t-}^{(n)} \text{and }\epsilon < t \}, & t > 0,
     \end{cases}
    \end{equation*}
    and note that,  due to 
    \eqref{eq:grid}, 
    \begin{equation} 
    \hat A_t ^{(n)} := \int_0 ^t h_s^{(n)} dN_s^{(n)} = \sum_{ \tau_k \leq t} \tau_{k} - \tau_{k-1} = \max\{ \tau_k \leq t\}
    \end{equation}
    defines a process that satisfies \eqref{eq:ConsistentHazard}, i.e. 
    $\lim _{n \rightarrow \infty} \sup_{t \leq \mathcal{T}} \big|  { \hat A}_t^{(n)}
    -t \big|  = 0,$ and Theorem \ref{theorem:ODE} holds.% It is also straight forward to see that $\{ \hat{A}^{(n)} \}_n$ is P-UT, so Theorem \ref{theorem:ODE} applies.   
\end{remark}

If $\sqrt n ( \pmb {\hat A}^{(n)}_t - \pmb A_t)$ defines a local martingale
that satisfies the martingale central limit theorem, 
then the sequence $\{\sqrt n ( \pmb {\hat X}^{(n)}_t - \pmb X_t)\}$ converges to a solution of a given SDE that is driven by a Gaussian process. This is the content of the following theorem. A proof can be found in the appendix.

\begin{theorem} \label{mainSDE}
Suppose that 
\begin{enumerate}[(i)]
  \item
   $F := \begin{pmatrix}   F_1, \dots, F_k \end{pmatrix}$ have bounded and
  continuous first and second derivatives on a domain that contains $\{\pmb X_t |t \in [0, \mathcal  T]  \}$,
  \item
  $ \pmb{ Z}_0^{(n)} := \sqrt n \big(  \pmb { \hat X} _0^{(n)} - \pmb X_0 \big)$ converges in law  to a mean zero Gaussian
  random vector $\pmb Z_0$ as $n \rightarrow \infty$,
  \item \label{enum:WPUT} $\pmb W^{(n)} := \sqrt n \big(  \pmb{\hat A}^{(n)} - \pmb A \big)$  %then we also assume that  
    %\begin{equation} 
   %\label{eq:pointwiseVariationConv}
    %              \lim_{n \rightarrow \infty} P \bigg( \sup_{t \leq T} \big| 
    %                 [\pmb{\hat W}^{(n)}]_t - [\pmb W]_t \big|\geq \epsilon
    %      \bigg)  = 0
     %     \end{equation}
  %or every $\epsilon > 0$, 
   %$\pmb W^{(n)}$ 
   converges in law (relative to the Skorohod-metric) towards a mean zero
   Gaussian martingale with independent increments and is P-UT
   (see
           Proposition \ref{prop:AalenPUT} for additive hazards),
   %satisfies the conditions of Lemma \ref{lemma:P-UTexplained},
   
   %\begin{equation} \label{eq:WnPUT}
   %   \sup_n E[ \sup_{t \leq \mathcal T}  \sqrt n | \pmb G^{(n)}_t|  ] < \infty.
   %\end{equation}
   \item \label{enum:brWPUT}
   $\{ [\pmb W^{(n)}] \}_n$ is P-UT
   (see
           Proposition \ref{prop:AalenPUT} for additive hazards), %satisfies the conditions of Lemma \ref{lemma:P-UTexplained}.
   
   %For each $t \in [0,\mathcal T]$,

   %\begin{equation} \label{eq:totvarquadr}
   %       \{ \sum_{i =1 }^l \int_0 ^t n \big|\pmb g_{s-}^{(n)i} \big|_2^2 d\pmb N_s^{(n)i}  \} _n    
%  \end{equation}
%  defines a tight family of finite random variables.
\end{enumerate}

Let $\pmb Z$ denote the unique solution to 
        the stochastic differential equation 
\begin{equation} \label{eq:mainLimit}
       \pmb Z_t = \pmb Z_0 + \sum_{j=1} ^k \int_0 ^t \nabla F_j ( \pmb X_{s-}) \pmb Z_{s-} dA_s^j + \int_0^t F( \pmb X_{s-}) d\pmb W_s. 
\end{equation}
Now,   
$\pmb Z$ is a mean zero Gaussian process, and
  $$\pmb Z^{(n)} := \sqrt n \big(  \pmb {\hat X}^{(n)} - \pmb X \big) $$ 
        converges in law (relative to the Skorohod-metric) to  $\pmb Z$ as $n \rightarrow \infty$.
        Moreover, let  $\pmb V_t$ denote the covariance for $\pmb Z_t$,
     suppose that $\{ \pmb{\hat V}_0^{(n)} \}$ is a consistent estimator for  $\pmb V_0$, and let $\pmb {\hat V}^{(n)}$ denote the solution of the stochastic differential equation: 
        \begin{equation}
        \begin{split}
                \pmb {\hat V}_t^{(n)}  = &\pmb {\hat V}_0^{(n)}  + \sum_{j = 1}^k \int_0^t \pmb
                {\hat V}_{s-}^{(n)}  \nabla F_j(\pmb {\hat X}_{s-}^{(n)} )^\intercal 
                + \nabla F_j(\pmb  {\hat X}_{s-}^{(n)} ) \pmb  {\hat V}_{s-}^{(n)}  d
                {\hat A}_{s}^{(n)j} \\ &
                + n \int_0 ^t F(\pmb { \hat X}_{s-}^{(n)} ) d{ [ \pmb  {\hat A}^{(n)} 
                        ]}_{s} F(\pmb  {\hat X}_{s-}^{(n)} )^\intercal.
                        \end{split}
                        \label{eq:XVarEst}
        \end{equation}
        Now, $\pmb {\hat V}^{(n)}$ defines a consistent estimator of $\pmb V$, i.e. 
        \begin{equation} \label{eq:consistentVarianceEst}
                \lim_{n \rightarrow \infty}P \bigg( \sup_{s \leq \mathcal T} \big|
          \pmb {\hat V}_s^{(n)}  -  \pmb V_s 
        \big|   \geq \epsilon \bigg) = 0 
        \end{equation}
        for every $\epsilon > 0$, and $\pmb V$ solves the following ordinary differential equation: 
                \begin{equation} \label{eq:contvar}
                \begin{split}
                \pmb V_t = & \pmb V_0 + \sum_{j = 1}^k \int_0^t \pmb V_s \nabla F_j(\pmb X_s)^\intercal 
                + \nabla F_j(\pmb X_s) \pmb V_s dA_s^j \\ &+ \int_0 ^t F(\pmb X_s) d [ \pmb W
                ]_s F(\pmb X_s)^\intercal.
                \end{split}
        \end{equation}

\end{theorem}

%\begin{lemma}
        %$Z_t$ is a mean zero Gaussian process and the covariance at time $t$ 
       %$V_t := E[Z_t
        %Z_t^\intercal]$ is given by the following ordinary differential
        %equation: 
        %\begin{equation}
                %V_t = V_0 + \sum_{j = 1}^k \int_0^t V_s \nabla F(X_s)^\intercal 
                %+ \nabla F(X_s) V_s dA_s^j + \int_0 ^t F(X_s) d\langle W
                %\rangle_s F(X_s)^\intercal 
        %\end{equation}
        %\begin{proof}
                %Note that $\langle W \rangle$ is deterministic since $W$ has
                %independent increments. 
                %\begin{align*}
                        %E\big[Z_t Z_t^\intercal\big] = &
                        %E\big[Z_0 Z_0^\intercal \big]
                        %+ E\bigg[\int_0^t Z_{s-} dZ_s^\intercal \bigg]
                        %+ E\bigg[\int_0^tdZ_s  Z_{s-}^\intercal \bigg]
                        %+ E\big[ [ Z, Z]_t\big] \\
                        %= &  E\big[Z_0 Z_0^\intercal \big]
                        %+  E\bigg[\sum_{j=1}^k \int_0^t Z_{s-} dA_s^j
                        %Z_{s-}^\intercal  \nabla F(X_s)^\intercal                         
                         %\bigg]
%\\&    + E\bigg[\sum_{j= 1}^k 
%\int_0^t  \nabla F(X_s) Z_{s-}dA^j _s Z_{s-}^\intercal           \bigg]
%\\ & + E\bigg[ \int_0^t F(X_{s-}) d [ W] _s
                        %F(X_{s-})^\intercal \bigg] \\
%= &  E\big[Z_0 Z_0^\intercal \big]
                        %+  \sum_{j=1}^k \int_0^tE\big[ Z_{s-}  Z_{s-}^\intercal
                        %\big]                         \nabla F(X_s)^\intercal   
                        %dA_s^j
%\\&    + \sum_{j= 1}^k 
%\int_0^t  \nabla F(X_s) E\big[Z_{s-} Z_{s-}^\intercal \big]      dA^j  
%\\ & + \int_0^t F(X_{s-}) d \langle W \rangle _s
                        %F(X_{s-})^\intercal.  
                %\end{align*}
        %\end{proof}
%\end{lemma}

\begin{remark}\label{remark:lebesgue variance argument}
	When $A_t = t$, we assume that the mesh introduced by the jump times $\{ \tau^{(n)}\}$ of $\hat A^{(n)}$ converge sufficiently fast to zero so that the limit of $\sqrt{n}(\hat{A}_t^{(n)} - A_t)$ is zero. This will e.g. hold if $\{ \tau^{(n)}\}$ is an equidistant partition, so that $\max_{k} | \tau_{k+1}^{(n)} - \tau_{k}^{(n)} | \leq \frac{\mathcal{T}}{n}$.  In that case, $\sqrt{n}(\hat{A}_t^{(n)} - A_t) = \sqrt{n}(\max\{\tau_{k} \leq t\} - t) \leq \sqrt{n}\max_{k} | \tau_{k+1}^{(n)} - \tau_{k}^{(n)} | \leq \frac{\mathcal{T}}{\sqrt{n}}$. In particular, this assumption will make the last integral in \eqref{eq:contvar} zero.
\end{remark}

 We can compute covariances and therefore also point-wise confidence  intervals for our estimates based on the ODEs from Section \ref{theorem:ODE}, by plugging in $\pmb {\hat X}^{(n)}$ for $\pmb{X}$ and $\pmb{  \hat A}^{(n)}$ for $\pmb A$. Note that, analogously to \eqref{eq:SDEpluginEst}, it is straight forward to solve the stochastic differential equation  \eqref{eq:XVarEst}, since it is driven by jump-processes. 
If $\tau_1, \tau_2,  \dots $ denote the jump times of  $\pmb {\hat A}^{(n)}$, then we have that 
\begin{align}
\begin{split}
        \pmb {\hat V}_t^{(n)} = & \pmb {\hat V}_{\tau_{k-1}}^{(n)} + 
        \sum_{j =1}^k \bigg( \pmb {\hat V}_{\tau_{k-1}}^{(n)} 
        \nabla F_j (  \pmb {\hat X}_{\tau_{k-1}}^{(n)}   )^{\intercal}
        +
        \nabla F_j (  \pmb {\hat X}_{\tau_{k-1}}^{(n)}   )
        \pmb {\hat V}_{\tau_{k-1}}^{(n)} 
        \bigg) 
        \Delta  \hat A_{\tau_{k}}^{(n)j}  \\ & + 
        n F (  \pmb {\hat X}_{\tau_{k-1}}^{(n)}   ) 
        %\Delta  \pmb {\hat A}_{\tau_{k}}^{(n)} 
        %\Delta  \pmb {\hat A}_{\tau_{k}}^{(n) \intercal} 
        \pmb H^{(n)}_{\tau_k}
        F (  \pmb {\hat X}_{\tau_{k-1}}^{(n)}   ) ^{\intercal}
       \end{split}
       \label{eq:VarianceEstimator}
\end{align}
with $k\times k$ matrix $\pmb H^{(n)}_t = [\pmb {\hat A}^{(n)} - \pmb A]_t$, whenever $\tau_{k} \leq t < \tau_{k+1}$. Following Remark \ref{remark:lebesgue variance argument} we have that $\big(\pmb H^{(n)}_t\big)_{i,j} = 0$ if either $A_t^i = t$ or $A_t^j = t$. Otherwise, $ \big(\pmb H^{(n)}_t \big)_{i,j} = \Delta \hat A_t^{(n),i} \Delta \hat A_t^{(n),j}$.

The properties $(\ref{enum:SDETightness})$ of Theorem \ref{theorem:ODE} and
$(\ref{enum:WPUT})$ and $(\ref{enum:brWPUT})$ of Theorem \ref{mainSDE}
are indeed satisfied for Aalens additive regression under
relatively weak assumptions. 
\begin{proposition} \label{prop:AalenPUT}
        Assume that we have $n$ i.i.d individuals and that the intensity of
        each $N^i$, at time $t$, is on the form
        $\lambda_t^i = Y^i_t \alpha_t^\intercal U^i_{t-}$, 
        where $\alpha$ is bounded and continuous. Write $\pmb{U}^{(n)}$ for the matrix whose $i$'th row is equal to $U^i$, and $\pmb{Y}^{(n)}$ for the diagonal matrix with $i$'th diagonal element is equal to $Y^i$. Suppose that 
        \begin{enumerate}
                \item $E[ \sup_{t \leq \mathcal{T}} |U_t^i|_3^3    ] < \infty$ for each $i$, 
                \item 
                        \begin{equation*}
                                \lim_{ J \rightarrow \infty} \inf_n P \bigg(
                        \sup_{t \leq \mathcal{T}} \Tr\bigg(  \Big( \frac {\pmb U^{(n)\intercal}_{t-}
                                \pmb Y^{(n)}_t
                                \pmb U_{t-}^{(n)}  }{ n}\Big) ^{-1}   \bigg) \geq J \bigg) = 0.
                \end{equation*}
        \end{enumerate}
        Then 
        \begin{equation}
	\label{eq:cumulative hazard estimator} \pmb A_t ^{(n)} := \int_0^t
        (\pmb U_{s-}^{(n)\intercal }  \pmb Y^{(n)}_{s} \pmb U_{s-}^{(n)})^{-1}
        \pmb U_{s-}^{(n)\intercal}  \pmb Y^{(n)}_{s} d \pmb N_s^{(n)}. 
\end{equation}
        satisfies $(\ref{enum:SDETightness})$ of Theorem \ref{theorem:ODE} and
$(\ref{enum:WPUT}$) and  $(\ref{enum:brWPUT}$) of Theorem \ref{mainSDE}.
\end{proposition}

\section{Plugin estimator examples}
\begin{figure}[ht]
    \centering
  \includegraphics[width=0.9\textwidth]{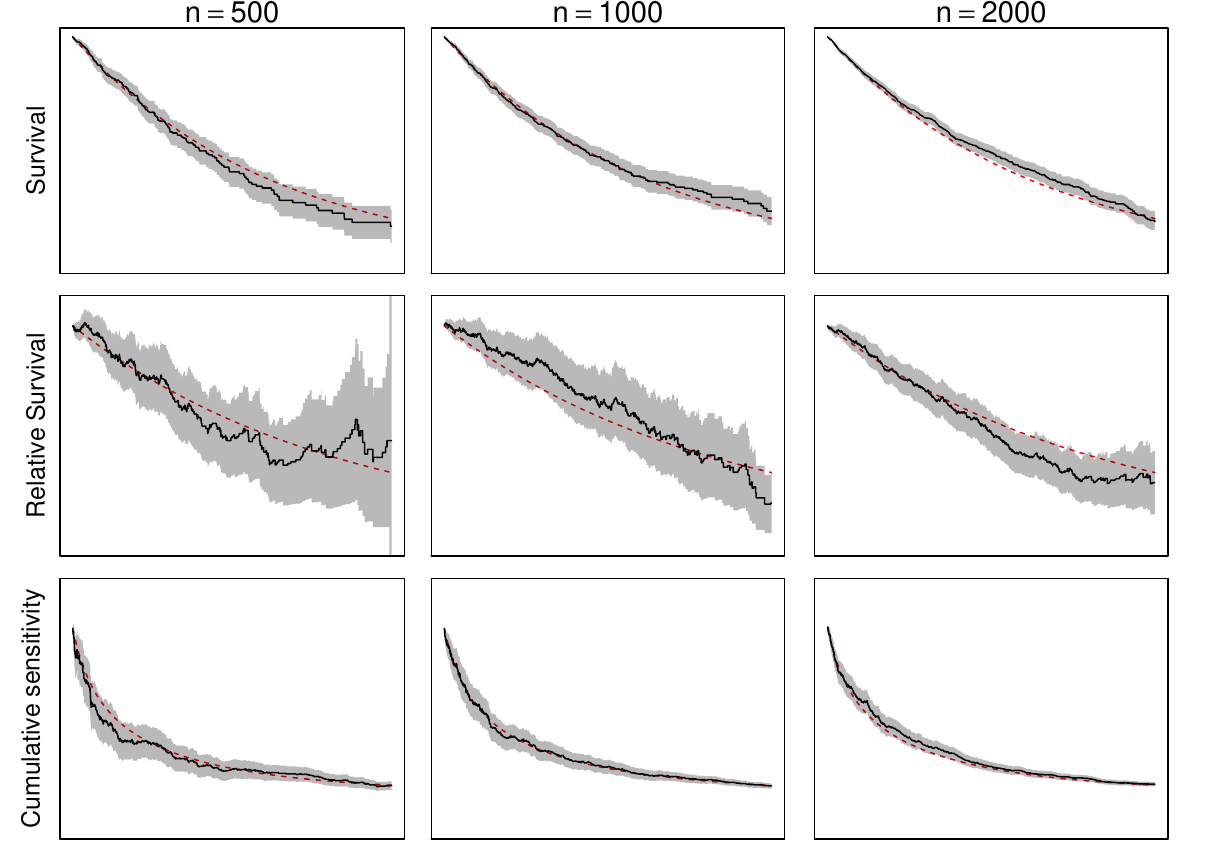}
    \caption{Plugin estimates for selected parameters with population sizes $500,1000,$ and $2000,$ along with 95 \% pointwise confidence intervals. Exact parameters is shown in dotted lines.}
    \label{fig:panelPlotParameters1}
\end{figure}

\begin{figure}[ht]
    \centering
  \includegraphics[width=0.9\textwidth]{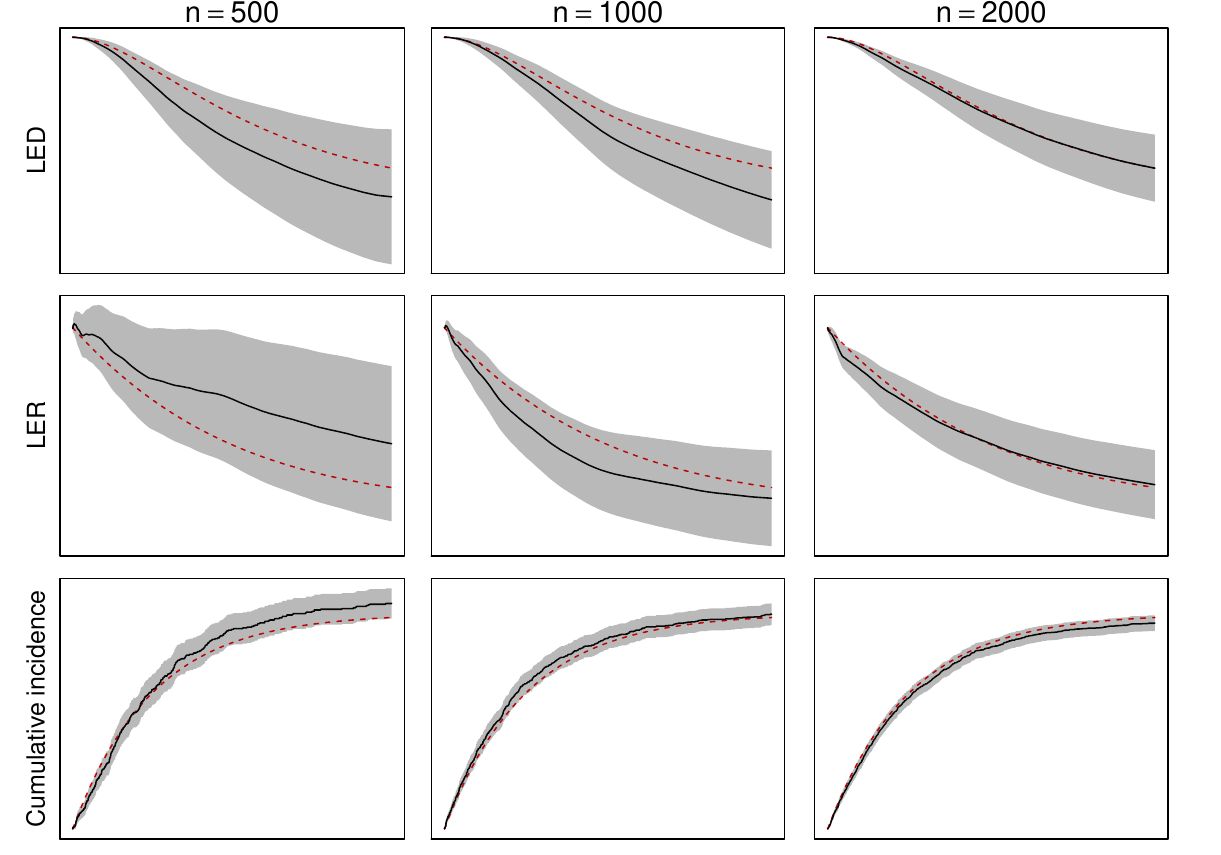}
    \caption{Plugin estimates for selected parameters with population sizes $500,1000,$ and $2000,$ along with 95 \% pointwise confidence intervals. Exact parameters is shown in dotted lines. LED and LER are abbreviations for life expectancy difference and life expectancy ratio, respectively.}
    \label{fig:panelPlotParameters2}
\end{figure}

In the following we write out the plugin estimators for some of the examples we showed in Section \ref{section:parameters}, using the notation introduced in Definition \ref{definition:SDEPluginEstimator}. The most involved variance expressions are moved to the Appendix. We will write $\Delta \tau_k := \tau_k - \tau_{k-1}$, and use $e_{i,j}$ to denote matrices with entries equal to zero in all positions apart from the $i$th row and $j$th column where the value is $1$, in order to keep the notation simple. Simulations of selected parameters are displayed in Figures \ref{fig:panelPlotParameters1} and \ref{fig:panelPlotParameters2}.
\subsection{Survival curves}
The survival plugin estimator is
\begin{align*}
    \hat{S}_t &= \hat{S}_{\tau_{k-1}} - \hat{S}_{\tau_{k-1}} \Delta\hat{A}_{\tau_k},
\end{align*}
which is nothing but the Kaplan-Meier estimator expressed as a difference equation. The variance plugin estimator is
\begin{align*}
    \hat{V}_t &= \hat{V}_{\tau_k-1} - 2\hat{V}_{\tau_k-1} \Delta \hat{A}_{\tau_k} + n\hat{S}_{\tau_{k-1}} (\Delta \hat{A}_{\tau_{k}})^2.
\end{align*}

\subsection{Relative survival}
The relative survival difference equation is
\begin{align*}
        \widehat{RS}_t %\frac{\hat S^1_t}{\hat S^0_t} 
        =  \widehat{RS}_{\tau_{k-1}} %\frac{\hat S^1_{\tau_{k-1}}}{\hat S^0_{\tau_{k-1}}} 
        +    
        \begin{pmatrix} -\widehat{RS}_{\tau_{k-1}} &
        \widehat{RS}_{\tau_{k-1}}\end{pmatrix}
        \Delta\begin{pmatrix} \hat{A}_{\tau_{k}}^1 \\ \hat{A}_{\tau_{k}}^0 \end{pmatrix}.
\end{align*}

The plugin variance solves
\begin{align*}
    \hat V_t = \hat V_{\tau_{k-1}} &-  2 \hat V_{\tau_{k-1}}
     \Delta \hat{A}_{\tau_k}^1 + 2\hat V_{\tau_{k-1}}
    \Delta \hat{A}_{\tau_k}^0 \\
    &+ n\begin{pmatrix} -\widehat{RS}_{\tau_{k-1}}%\frac{\hat S^1_{\tau_{k-1}}}{\hat S^0_{\tau_{k-1}}} 
    & \widehat{RS}_{\tau_{k-1}} %\frac{\hat S^1_{\tau_{k-1}}}{\hat S^0_{\tau_{k-1}}}  
    \end{pmatrix} \Delta \begin{pmatrix}
        \hat A_{\tau_{k}}^1 \\
        \hat A_{\tau_{k}}^0
    \end{pmatrix}
    \Delta \begin{pmatrix}
        \hat A_{\tau_{k}}^1 & \hat A_{\tau_{k}}^0
    \end{pmatrix}
    \begin{pmatrix} 
        -\widehat{RS}_{\tau_{k-1}}%\frac{\hat S^1_{\tau_{k-1}}}{\hat S^0_{\tau_{k-1}}} 
        \\\widehat{RS}_{\tau_{k-1}}%\frac{\hat S^1_{\tau_{k-1}}}{\hat S^0_{\tau_{k-1}}}
    \end{pmatrix}.
\end{align*}

\subsection{Mean and restricted mean survival}
The restricted mean survival plugin estimator solves
\begin{align*}
    \begin{pmatrix}
    \hat{R}_{t} \\
    \hat{S}_{t}
    \end{pmatrix} &=
    \begin{pmatrix}
    \hat{R}_{\tau_{k-1}} \\
    \hat{S}_{\tau_{k-1}}
    \end{pmatrix}+
    \begin{pmatrix}
    \hat{S}_{\tau_{k-1}} & 0 \\
    0 & -\hat{S}_{\tau_{k-1}}
    \end{pmatrix}
    \Delta
     \begin{pmatrix}
    \tau_{k} \\
    \hat{A}_{\tau_k}
    \end{pmatrix}.
\end{align*}
The variance expression reads
\begin{align*}
    \pmb{\hat V}_t &= \pmb{\hat V}_{\tau_{k-1}} + \Big( \pmb{\hat V}_{\tau_{k-1}}e_{2,1} %\begin{pmatrix} 0&0\\1&0\end{pmatrix} + \begin{pmatrix} 0&1\\0&0\end{pmatrix}
    +e_{1,2}\pmb{\hat V}_{\tau_{k-1}} \Big) \Delta \tau_{k} \\
    &- \Big( \pmb{\hat V}_{\tau_{k-1}} e_{2,2}%\begin{pmatrix} 0&0\\0&1\end{pmatrix} + \begin{pmatrix} 0&0\\0&1\end{pmatrix}
    +e_{2,2}\pmb{\hat V}_{\tau_{k-1}} \Big) \Delta \hat{A}_{\tau_k} \\
    &+ n\begin{pmatrix} \hat{S}_{\tau_{k-1}}&0\\0&-\hat{S}_{\tau_{k-1}}\end{pmatrix} \Delta\begin{pmatrix} 0\\\hat{A}_{\tau_k} \end{pmatrix}\Delta\begin{pmatrix} 0 & \hat{A}_{\tau_k} \end{pmatrix} \begin{pmatrix} \hat{S}_{\tau_{k-1}}&0\\0&-\hat{S}_{\tau_{k-1}}\end{pmatrix}.
\end{align*}

\subsection{Life expectancy difference and life expectancy ratio}
The LED plugin estimator is
\begin{align*}
    \begin{pmatrix}
    \widehat{\text{LED}}_{t}^{12} \\
    \hat{S}_{t}^1 \\
    \hat{S}_{t}^2
    \end{pmatrix} &=
    \begin{pmatrix}
    \widehat{\text{LED}}_{\tau_{k-1}}^{12} \\
    \hat{S}_{\tau_{k-1}}^1 \\
    \hat{S}_{\tau_{k-1}}^2
    \end{pmatrix}+
    \begin{pmatrix}
    \hat{S}_{\tau_{k-1}}^1 - \hat{S}_{\tau_{k-1}}^2 & 0 & 0  \\
    0 & -\hat{S}_{\tau_{k-1}}^1 & 0  \\
    0 & 0 & -\hat{S}_{\tau_{k-1}}^2 
    \end{pmatrix}
    \Delta
    \begin{pmatrix}
    \tau_k\\
    \hat{A}_{\tau_{k}}^1\\
    \hat{A}_{\tau_{k}}^2
    \end{pmatrix},
\end{align*}

with variance equation

\begin{align*}
    \pmb{\hat V}_t &= \pmb{\hat V}_{\tau_{k-1}} + \Big( \pmb{\hat V}_{\tau_{k-1}}(e_{2,1}-e_{3,1}) %\begin{pmatrix} 0&0&0\\1&0&0\\-1&0&0\end{pmatrix} + \begin{pmatrix} 0&1&-1\\0&0&0\\0&0&0\end{pmatrix}
    + (e_{1,2}-e_{1,3})\pmb{\hat V}_{\tau_{k-1}} \Big) \Delta \tau_{k} \\
    &- \Big( \pmb{\hat V}_{\tau_{k-1}} e_{2,2}%\begin{pmatrix} 0&0&0\\0&-1&0\\0&0&0\end{pmatrix} + \begin{pmatrix} 0&0&0\\0&-1&0\\0&0&0\end{pmatrix}
    +e_{2,2}\pmb{\hat V}_{\tau_{k-1}} \Big) \Delta \hat{A}_{\tau_k}^1 \\
    &- \Big( \pmb{\hat V}_{\tau_{k-1}} e_{3,3}%\begin{pmatrix} 0&0&0\\0&0&0\\0&0&-1\end{pmatrix} + \begin{pmatrix} 0&0&0\\0&0&0\\0&0&-1\end{pmatrix}
    + e_{3,3} \pmb{\hat V}_{\tau_{k-1}} \Big) \Delta \hat{A}_{\tau_k}^2 \\
    & + nF_{\tau_{k-1}} \Delta\begin{pmatrix} 0\\\hat{A}_{\tau_k}^1\\\hat{A}_{\tau_k}^2 \end{pmatrix}\Delta\begin{pmatrix} 0 & \hat{A}_{\tau_k}^1 & \hat{A}_{\tau_k}^2 \end{pmatrix} F_{\tau_{k-1}}^\top.
\end{align*}
Here $F$ is the LED integrand function, i.e.\
\begin{align*}
    F_{\tau_{k-1}} &=
    \begin{pmatrix} \hat{S}_{\tau_{k-1}}^1-\hat{S}_{\tau_{k-1}}^2&0&0\\ 0&\hat{S}_{\tau_{k-1}}^1&0\\0&0&\hat{S}_{\tau_{k-1}}^2\end{pmatrix}.
\end{align*}

The LER plugin estimator reads
\begin{align*}
    \begin{pmatrix}
        \widehat{\text{LER}}_{t}^{12} \\
        \hat{S}_{t}^1 \\
        \hat{S}_{t}^2 \\
        \hat{R}_t^1 \\
        \hat{R}_t^2
    \end{pmatrix} &=
    \begin{pmatrix}
        \widehat{\text{LER}}_{\tau_{k-1}}^{12} \\
        \hat{S}_{\tau_{k-1}}^1 \\
        \hat{S}_{\tau_{k-1}}^2 \\
        \hat{R}_{\tau_{k-1}}^1 \\
        \hat{R}_{\tau_{k-1}}^2
    \end{pmatrix}+
    \begin{pmatrix}
        \frac{\hat{S}_{\tau_{k-1}}^1 \hat R_{\tau_{k-1}}^2 - \hat{S}_{\tau_{k-1}}^2 \hat R_{\tau_{k-1}}^1}{(\hat R_{\tau_{k-1}}^2)^2} & 0 & 0  \\
        0 & -\hat{S}_{\tau_{k-1}}^1 & 0  \\
        0 & 0 & -\hat{S}_{\tau_{k-1}}^2  \\
        \hat{S}_{\tau_{k-1}}^1 & 0 & 0 \\
        \hat{S}_{\tau_{k-1}}^2 & 0 & 0
    \end{pmatrix}
    \Delta
    \begin{pmatrix}
        \tau_k \\
        \hat{A}_{\tau_{k}}^1\\
        \hat{A}_{\tau_{k}}^2
    \end{pmatrix}.
\end{align*}

The variance expression is involved, and can be found in the Appedix.

\subsection{Cumulative incidence and competing risks}
The plugin estimator for the cumulative incidence example is 
\begin{align*}
        \begin{pmatrix}
                \hat{S}_t \\ 
                \hat{C}^1_t \\ \vdots \\\hat{C}^m_t
        \end{pmatrix}
        = \begin{pmatrix}
                \hat{S}_{\tau_{k-1}} \\\hat{C}^1_{\tau_{k-1}} \\\vdots\\ \hat{C}^m_{\tau_{k-1}}
        \end{pmatrix}
        +          
        \begin{pmatrix}
        -\hat{S}_{\tau_{k-1}} & -\hat{S}_{\tau_{k-1}}  & \dots & -\hat{S}_{\tau_{k-1}} \\
        \hat{S}_{\tau_{k-1}}  & 0 & \dots & 0\\
        0 & \dots & \ddots  & \hat{S}_{\tau_{k-1}}
        \end{pmatrix}
        \Delta \begin{pmatrix}
         \hat{A}^1_{\tau_{k}} \\ \vdots \\ \hat{A}^m_{\tau_{k}}
        \end{pmatrix},
\end{align*}
The plugin variance equation is
\begin{align*}
    \pmb{\hat V}_t &= \pmb{\hat V}_{\tau_{k-1}} %+ \Big( \pmb{\hat V}_{\tau_{k-1}} G_{\tau_{k-1}}^\top + G_{\tau_{k-1}}\pmb{\hat V}_{\tau_{k-1}} \Big) \Delta \tau_{k} \\
    + \Big( \pmb{\hat V}_{\tau_{k-1}} (e_{1,2}-e_{1,1})% begin{pmatrix} -1&1&0&\cdots&0\\0&0&0&\cdots&0\\0&0&0&\cdots&0\\\vdots&\vdots&\vdots&\ddots&\vdots\\0&0&0&\cdots&0\end{pmatrix} + \begin{pmatrix} -1&0&0&\cdots&0\\1&0&0&\cdots&0\\0&0&0&\cdots&0\\\vdots&\vdots&\vdots&\ddots&\vdots\\0&0&0&\cdots&0\end{pmatrix}
    + (e_{2,1}-e_{1,1})\pmb{\hat V}_{\tau_{k-1}} \Big) \Delta \hat{A}_{\tau_k}^1 \\
    &+ \Big( \pmb{\hat V}_{\tau_{k-1}} (e_{1,3}-e_{1,1})%\begin{pmatrix} -1&0&1&\cdots&0\\0&0&0&\cdots&0\\0&0&0&\cdots&0\\\vdots&\vdots&\vdots&\ddots&\vdots\\0&0&0&\cdots&0\end{pmatrix} + \begin{pmatrix} -1&0&0&\cdots&0\\0&0&0&\cdots&0\\1&0&0&\cdots&0\\\vdots&\vdots&\vdots&\ddots&\vdots\\0&0&0&\cdots&0\end{pmatrix}
    +(e_{3,1}-e_{1,1})\pmb{\hat V}_{\tau_{k-1}} \Big) \Delta \hat{A}_{\tau_k}^2 \\
    & \vdots \\
    &+ \Big( \pmb{\hat V}_{\tau_{k-1}} (e_{1,m+1}-e_{1,1})%\begin{pmatrix} -1&0&0&\cdots&1\\0&0&0&\cdots&0\\0&0&0&\cdots&0\\\vdots&\vdots&\vdots&\ddots&\vdots\\0&0&0&\cdots&0\end{pmatrix} + \begin{pmatrix} -1&0&0&\cdots&0\\0&0&0&\cdots&0\\0&0&0&\cdots&0\\\vdots&\vdots&\vdots&\ddots&\vdots\\1&0&0&\cdots&0\end{pmatrix}
    +(e_{m+1,1}-e_{1,1})\pmb{\hat V}_{\tau_{k-1}} \Big) \Delta \hat{A}_{\tau_k}^m \\
    &+ nF_{\tau_{k-1}} \Delta\begin{pmatrix}
    \hat{A}_{\tau_k}^1\\\vdots\\\hat{A}_{\tau_k}^m \end{pmatrix}\Delta\begin{pmatrix} \hat{A}_{\tau_k}^1 & \cdots & \hat{A}_{\tau_k}^m \end{pmatrix} F_{\tau_{k-1}}^\top.
\end{align*}

\subsection{Mean frequency function for recurrent events}
We get a plugin estimator equation that reads
\begin{align*}
    \begin{pmatrix}
    \hat{K}_t \\
    \hat{S}_t
    \end{pmatrix}
    &= 
    \begin{pmatrix}
    \hat{K}_{\tau_{k-1}} \\
    \hat{S}_{\tau_{k-1}}
    \end{pmatrix}+
    \begin{pmatrix}
        \hat{S}_{\tau_{k-1}} & 0  \\
        0 & -\hat{S}_{\tau_{k-1}}
    \end{pmatrix}
    \Delta \begin{pmatrix}
        \hat{A}_{\tau_{k}}^H \\
        \hat{A}_{\tau_{k}}^D
    \end{pmatrix}.
\end{align*}
\normalsize
The variance is 
\begin{align*}
    \pmb{\hat V}_t &= \pmb{\hat V}_{\tau_{k-1}} + \Big( \pmb{\hat V}_{\tau_{k-1}}e_{2,1} %\begin{pmatrix} 0&0\\1&0 \end{pmatrix} + \begin{pmatrix} 0&1\\0&0 \end{pmatrix} 
    +e_{1,2}\pmb{\hat V}_{\tau_{k-1}} \Big) \Delta \hat A_{\tau_k}^H \\
    &- \Big( \pmb{\hat V}_{\tau_{k-1}} e_{2,2}%\begin{pmatrix} 0&0\\0&-1 \end{pmatrix} + \begin{pmatrix} 0&0\\0&-1 \end{pmatrix} 
    +e_{2,2}\pmb{\hat V}_{\tau_{k-1}} \Big) \Delta \hat A_{\tau_k}^D \\
    &+ n\begin{pmatrix}
        \hat{S}_{\tau_{k-1}} & 0  \\
        0 & -\hat{S}_{\tau_{k-1}}
    \end{pmatrix} \Delta\begin{pmatrix} \hat{A}_{\tau_k}^H\\\hat{A}_{\tau_k}^D \end{pmatrix}\Delta\begin{pmatrix} \hat{A}_{\tau_k}^H & \hat{A}_{\tau_k}^D \end{pmatrix} \begin{pmatrix}
        \hat{S}_{\tau_{k-1}} & 0  \\
        0 & -\hat{S}_{\tau_{k-1}}
    \end{pmatrix}.
\end{align*}

\subsection{Cumulative sensitivity and specificity after screening}
We have
\begin{align*}
        \begin{pmatrix}
                \hat U_t \\ \hat V_t \\ \hat W_t \\\hat{X}_t  
        \end{pmatrix}=
         \begin{pmatrix}
                \hat U_{\tau_{k-1}} \\\hat V_{\tau_{k-1}} \\ \hat W_{\tau_{k-1}} \\\hat X_{\tau_{k-1}}  
        \end{pmatrix}   +  \begin{pmatrix}
               1 -\hat U_{\tau_{k-1}}  & 0  \\
               0  & -\hat V_{\tau_{k-1}} \\
               \frac{\hat W^2_{\tau_{k-1}} ( 1- \hat V_{\tau_{k-1}})( 1-
                         \hat U_{\tau_{k-1}}) }{\gamma \hat U_{\tau_{k-1}}^2}  
               &
                - \frac{\hat W^2_{\tau_{k-1}} \hat V_{\tau_{k-1}}\hat U_{\tau_{k-1}} } { \gamma \hat U_{\tau_{k-1}}^2 } 
                \\ 
                \frac{ \gamma \hat X_{\tau_{k-1}}^2 (1-
            \hat    U_{\tau_{k-1}}) }{\hat V_{\tau_{k-1}} }  & -
                \frac{\gamma \hat X_{\tau_{k-1}}^2 (1-
                \hat U_{\tau_{k-1}}) }{\hat V_{\tau_{k-1}} }
        \end{pmatrix}
        \Delta \begin{pmatrix}
        	\hat{A}^1_{\tau_{k}} \\\hat{A}^0_{\tau_{k}}
        \end{pmatrix},
\end{align*}
The variance expression can be found in the Appendix.

\section{Performance}
\subsection{Convergence}
\label{subsection:convergence}
Our convergence assessment involves two steps. In step one we find close approximations to the parameters, using the notation $\tilde{X}$. In step two we calculate plugin estimators for a range of sample sizes. We then check how fast the plugin estimators converge to $\tilde{X}$ as the sample size increases.

For performing step one we have developed code that generates survival data for given input hazards. Using the hazards we can calculate $\tilde{X}$ to desired precision. For instance, knowing the hazard $\alpha$ we can obtain the survival $e^{-\int \alpha_s ds}$ by numerical integration. %In this way we find reference solutions that serve the role as 'exact' solutions, and label them $\tilde{X}$.

For performing step two we simulate data, enabling us to estimate cumulative hazards, and therefore parameters specified by \eqref{eq:SDEpluginEst}. This step is performed for a range of sample sizes. For each sample size $n$ we simulate $k$ populations, such that we obtain plugin estimates $\hat{X}^{n,j}$, for $j=1,\cdots,k$. We check convergence using the following $L^2$ criterion:
\begin{align}
    L(n) &= \frac{1}{k}\sum_{j=1}^k\int_0^\mathcal{T} ( \tilde{X}_s -  \hat{X}_s^{n,j}  )^2 ds. \label{eq:convTerm}
\end{align}

\begin{figure}[ht]
    \centering
    \includegraphics[width=0.9\textwidth]{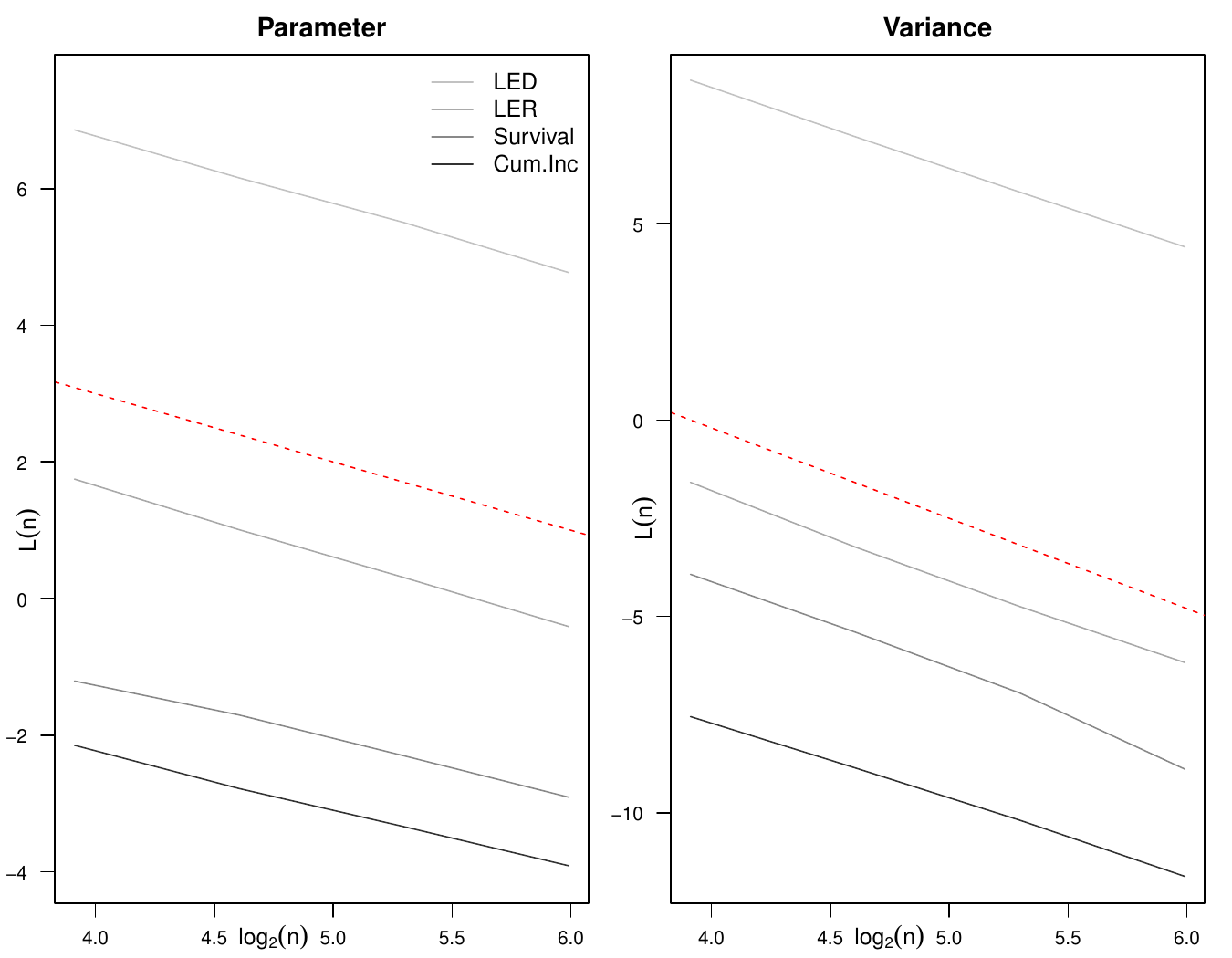}
    \caption{Convergence of selected plugin estimators; parameter to the left and variances to the right. We used dashed lines indicating convergence order $1$ in the left panel and $2.3$ in the right panel. LED and LER are abbreviations for life expectancy difference and life expectancy ratio, respectively.}
    \label{fig:jointConvConst}
\end{figure}

For assessing convergence of the plugin variance estimators we first obtain a variance estimate from a large bootstrap sample. As in step two above, we simulate data and obtain $k$ plugin variance estimates for a range of sample sizes $n$. In this way the same criterion \eqref{eq:convTerm} can be used.
%Convergence of the variance estimator is assessed by comparing with bootstrapped variance. Using a large bootstrap sample we can obtain a reference variance $\tilde{\sigma}^2$ for the parameter. As above we simulate trajectories $\hat{\sigma}^{2,n,j}$, and check convergence using the same criterion \eqref{eq:convTerm}.
Convergence plots are shown in Figure \ref{fig:jointConvConst}. %We added dashed lines indicating order of convergence.

%\textbf{Remark 1:} Theoretical results regarding the convergence can be derived under assumptions on \eqref{eq:Wn} and $E\Big[\big| \pmb X_0 - \pmb X_0^n \big|^2\Big]$. An application of Grönwall's inequality gives
%\begin{align*}
% \int_0^\mathcal{T} E\Big[ \big| \pmb X_t - \pmb X_t^n \big|^2  \Big] dt \leq \int_0^\mathcal{T} g_t dt + \int_0^\mathcal{T} \int_0^t g_s \alpha_s^2 e^{\int_s^t \alpha_u^2 du} ds dt,
%\end{align*}
%where
%\begin{align*}
%	g_t &= \frac{K_1}{n} E \Big[ \big| \int_0^t F(\pmb X_s^n) d\pmb W_s^n \big|^2 \Big] + K_1E\Big[ \big| \pmb X_0 - \pmb X_0^n \big|^2 \Big],
%\end{align*}
%for some constant $K_1$. If we, for instance, assume that 
%\begin{align*}
	%\sup_{t \in [0,\mathcal{T}]} E \Big[ \big| \int_0^t F(\pmb X_s^n) d\pmb W_s^n \big|^2 \Big]
%\end{align*}
%is uniformly bounded in $n$, and 
%\begin{align*}
	%E\Big[ \big| \pmb X_0 - \pmb X_0^n \big|^2 \Big] \leq \frac{K_2}{n},
%\end{align*}
%for some constant $K_2$, we would have a theoretical convergence of order one.

%%%%%
Figure \ref{fig:jointConvConst} suggests that the LED estimator performs worse than the other estimators. This happens because the variance depends on the time scale, since we are approximating Lebesgue integrals. The LER plugin variance also depends on the time scale, but here the time contribution is smaller.
%%%%%

%\begin{figure}[ht]
%    \centering
%    \includegraphics[width=\textwidth]{paramConvConst}
%    \caption{Convergence of selected plugin parameter estimates. Order $1$ is shown for reference. }
%    \label{fig:parameterConvergence}
%\end{figure}

\subsection{Coverage}

To estimate coverage we first obtain close approximations to the parameters, $\tilde{X}$, as described in step one in Section \ref{subsection:convergence}. Next, we simulated data to calculate plugin estimates, which were used to obtain confidence intervals. This step is repeated until we have a large collection of confidence intervals. At time $t$, a Bernoulli trial decides whether a confidence interval covers $\tilde{X}_t$. We estimate the expected coverage at time $t$ by this Bernoulli probability, i.e. by the average number of confidence intervals that cover $\tilde{X}_t$. 

We calculate coverage for two scenarios; constant hazards, and linear (crossing) hazards over a time period $[0,\mathcal{T}]$. We selected linear hazards that were large initially but linearly decreasing, or small initially but linearly increasing, such that they crossed each other at the halfway point $\mathcal{T}/2$.

\begin{figure}[ht]
    \centering
   	\includegraphics[width=0.9\textwidth]{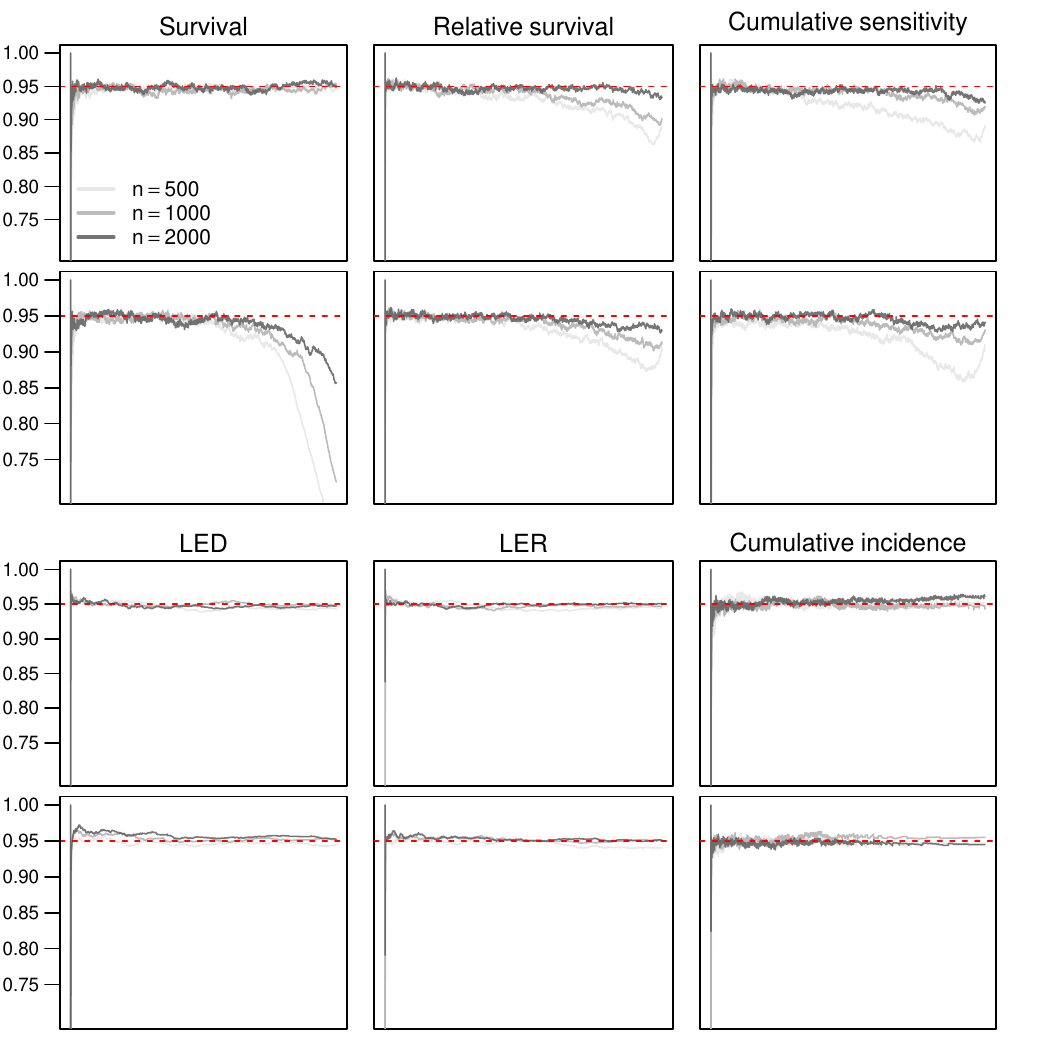}
    \caption{Estimated mean coverage for selected parameters simulated with constant(upper panel) and linearly crossing(lower panel) hazards. The dotted line indicates the confidence level. LED and LER are abbreviations for life expectancy difference and life expectancy ratio, respectively. }
    \label{fig:coverage}
\end{figure}

Coverage is shown in Figure \ref{fig:coverage}, suggesting that the coverage drops below the confidence level in some scenarios. This behavior is due to the plugin variance estimators. The Survival plugin variance, for instance, will drop drastically if there are events when few people are at risk, causing the Survival confidence interval to be narrow thereafter. Incidentally, the Greenwood estimator would give similar performance, since it is small whenever the Kaplan--Meier curve is small.

Overall the plugin estimators behave satisfactory, not only when the hazards are constant but also when the hazards are linearly crossing.

%section{Discussion}
%he concept of hazards is fundamental to survival analysis, and semi-parametric hazard models allow for simple adjustments of covariates. However, effect estimates on the hazard scale are hard to interpret. To obtain alternative effect measures, we have developed an approach that transforms cumulative hazard estimates to other scales. The transformations are derived by expressing effect measures as solutions of ordinary differential equations driven by cumulative hazards. We have used existing theory for stochastic differential equations to prove the consistency of the estimators and the variances. These derivations allows us to express plugin estimators for several effect measures, and our simulation studies suggest that the plugin estimates are well-behaved. 

%The concept of hazards is appealing because it allows for elegant, semi-parametric modeling. The concept of hazards has been fundamental to survival analysis, and semi-parametric hazard models allows for simple adjustments of covariates. However, estimates on hazard scale are hard to interpret causally. 

\section{Software}

\texttt{R} software for calculating plugin estimators, illustrated on simulated data examples, can be found on 

\noindent
\texttt{https://github.com/palryalen/transform.hazards}. For questions or comments regarding the shared code, contact p.c.ryalen@medisin.uio.no.

\section{Funding}
The authors were all supported by the research grant NFR239956/F20 - Analyzing clinical health registries: Improved software and mathematics of identifiability.

\section*{Appendix: Proofs}
\subsection{Proof of Lemma \ref{lemma:P-UTexplained}}

  \begin{proof}
                    We start by proving that $\{\pmb{M}^{(n)}\}_n$ is P-UT. Suppose that $\pmb{H}$ is predictable, matrix-valued and such that $\Tr (\pmb H_s^\intercal \pmb H_s) $ bounded by $1$. Note that for every optional stopping time $\tau$:  
                        \begin{align*}
                                 & E\big[ \big|\int_0^\tau \pmb H_s d \pmb M_s^{(n)} \big|_2^2
                                 \big] = 
                                E\big[ \text{Tr} \int_0^\tau \pmb H_s d\pmb M_s^{(n)} 
                                \big(\int_0^\tau \pmb H_s d \pmb M_s^{(n)} \big)^\intercal \big] \\ 
                                   = & E\big[ \text{Tr} \int_0^\tau \pmb H_s  d\langle
                                  \pmb M^{(n)} \rangle_s 
                                 \pmb H_s^\intercal \big]  \leq 
                                 E\big[\int_0^\tau \Tr (\pmb H_s^\intercal \pmb H_s)  d  \text{Tr} \langle
                                 \pmb M^{(n)} \rangle_s 
                                  \big]  
                                 \\
                                  \leq &  E\big[\int_0^\tau d  \text{Tr} \langle
                                 \pmb M^{(n)} \rangle_s 
                                  \big]  =  
                                  E\big[ \text{Tr} \langle
                                 \pmb M^{(n)} \rangle_\tau 
                                  \big]. 
                        \end{align*}
                        The  Lenglart-inequality \cite[ I-3.31]{JacodShiryaev}
                        therefore implies that  $\{\pmb M^{(n)}\}_n$ 
                        is P-UT.

                        For the first term $\ref{eq:tightRho})$ we have that $\text{Var}(\int_0^\cdot
                        \pmb \rho_s^{(n)} ds)_t= \int_0^t |\pmb \rho_s^{(n)} |_1ds $ is uniformly tight for every $t$, which by \cite[ VI 6.12]{JacodShiryaev} implies that $\{\pmb \rho^{(n)}\}_n$ is P-UT. Finally, the claim follows since $\{\pmb Z^{(n)}\}_n$ is the sum of two sequences that are P-UT.

            \end{proof}

\subsection{Proof of Theorem \ref{theorem:ODE}}
  \begin{proof}
  
We are given a series of SDEs that converges to an SDE (actually just an ODE),
and want to assure that this means that their solutions also converge to the
solution of the limit equation. This is a claim about stability of
stochastic differential equations. We want to prove our claim with an application
of \cite[Theorem IX.6.9]{JacodShiryaev}. Translated into our situation, it says
that if $
        (\pmb {\hat X}_0 ^{(n)}, \pmb {\hat A} ^{(n)} )$ converges in
        law to $(\pmb X_0, \pmb A)$ and the series 
         $\{\pmb {\hat A} ^{(n)} \}_n$ is predictably uniformly tight
         (P-UT), then $(\pmb {\hat X}_0 ^{(n)}, \pmb {\hat A} ^{(n)} ,\pmb {\hat X} ^{(n)} )$  also converges in law to $(\pmb X_0, \pmb A, \pmb X)$.
        
Moreover, \cite[VI.3.33]{JacodShiryaev} ensures convergence of  $(\pmb {\hat X}_0^{(n)}, \pmb {\hat A}^{(n)})$. The advertised result follows from the continuity of the projection mapping.%, when we recall that the $\mathbb{D}(\mathbb{R}^{n \times k \times n})$ Skorokhod topology is strictly finer than the product topology.
\end{proof}

\subsection{Proof of Theorem \ref{mainSDE}}

\begin{proof}

We will first prove that $\{\pmb {\hat Z}^{(n)}\}$ converges in law to $\pmb Z$
by another application of the stability result for SDEs found in \cite[Theorem IX.6.9]{JacodShiryaev}.
Then we will prove that the covariance  of $\pmb Z$ is given by the ODE \eqref{eq:contvar}.
Finally, we
will apply the SDE-stability result once more to see that the solutions of the SDE
\eqref{eq:XVarEst} converge in law to the covariance of $\pmb Z$.

To prove that   $\{\pmb Z^{(n)}\}$ converges in law, we start with the
following approximation of the derivative of $F_j$ in the direction $z$:
\begin{equation}
        DF^n_j  ( x, z):= \sqrt n \big(   F_j  ( x + n^{-1/2} z) - F_j ( x)
        \big),
\end{equation} 
let $        DF^n ( x,z) :=  
        \begin{pmatrix}
                 DF^n_1  ( x, z), &  \dots &  ,DF^n_l  ( x, z) \end{pmatrix}$ and
         note that we have $$\lim_n DF^n_j  (x, z) = \nabla F_j ( x) z.$$

Now,
\begin{align*}
        & \pmb {Z}_t^{(n)}  =   \sqrt n \big( \pmb {\hat X}_t^{(n)} -  \pmb X_t \big)\\  = & 
        \pmb { Z}_0^{(n)} 
       + 
        \sum_{j= 1}^k \int_0^t   \sqrt n \bigg( F_j ( \pmb {\hat
                X}_{s-}^{(n)}) - F_j ( \pmb X_s)
        \bigg)d {\hat A}_s
        ^{j(n)} 
         +   \int_0^t F ( \pmb X_s) d \pmb W^{(n)}_s  \\ = & 
         \pmb { Z}_0^{(n)} +  \sum_{j=1}^k \int_0^t DF^n_j  ( \pmb X_s, \pmb Z^{(n)}_{s-})
         d\hat A^{j(n)}_s 
          + \int_0^t F(\pmb X_s) d \pmb W^{(n)}_s
          \\ = &  
           \pmb { Z}_0^{(n)} + \int_0^t DF^n  ( \pmb X_s, \pmb Z^{(n)}_{s-})
         d\pmb {\hat A}^{(n)}_s 
          + \int_0^t F(\pmb X_s) d \pmb W^{(n)}_s. 
        \end{align*}

%Moreover,  \eqref{eq:WnPUT} and  \cite[Corollary VI.6.30]{JacodShiryaev} imply that
% $\{\pmb W^{(n)}\}_n$ is predictably uniformly tight. 
Now  \cite[Theorem VI.6.22]{JacodShiryaev} implies that 
$\{  \pmb { Z}_0^{(n)} +    \int_0^\cdot F(\pmb X_s) d \pmb W^{(n)}_s \}_n$ converges is law to 
$\pmb Z_0 + \int_0^\cdot  F(\pmb X_s) d \pmb W_s $, and since $\pmb A$ is continuous, \cite[Corollary VI.3.33]{JacodShiryaev} implies that 
$\{  (\pmb { Z}_0^{(n)} +    \int_0^\cdot  F(\pmb X_s) d \pmb W^{(n)}_s , \pmb{\hat A}^{(n)})\}_n$ converges in law to $(\pmb Z_0 + \int_0^\cdot  F(\pmb X_s) d \pmb W_s , \pmb A)$. 

Since $F$ has bounded and continuous first and second derivatives, a calculus
argument shows that $\{ DF^n \}_n$ have uniformly bounded  Lipschitz constants
by the Lipcschitz constant of $F$. 
This enables us to apply the stability result for SDEs found in \cite[Theorem
IX.6.9]{JacodShiryaev} again. It says that since $\{\pmb {\hat A}^{(n)}\}_n$ is
predictably uniformly tight, we also have that  
$\{  (\pmb { Z}_0^{(n)} +    \int_0^\cdot  F(\pmb X_s) d \pmb W^{(n)}_s , \pmb{\hat
        A}^{(n)}, \pmb Z^{(n)} )\}_n$ 
converges in law to 
$(\pmb Z_0 + \int_0^\cdot  F(\pmb X_s) d \pmb W_s , \pmb A, \pmb Z)$.

 To see that the covariance  of $\pmb Z$ is given by the ODE
 \eqref{eq:contvar}, we note that $[ \pmb W ]$ is deterministic since $\pmb W$ is a continuous Gaussian martingale
    with independent increments, and that  
                \begin{align*}
                        E\big[\pmb Z_t \pmb Z_t^\intercal\big] = &
                        E\big[\pmb Z_0 \pmb Z_0^\intercal \big]
                        + E\bigg[\int_0^t \pmb Z_{s-} d\pmb Z_s^\intercal \bigg]
                        + E\bigg[\int_0^td\pmb Z_s  \pmb Z_{s-}^\intercal \bigg]
                        + E\big[ [ \pmb Z, \pmb Z]_t\big] \\
                        = &  E\big[\pmb Z_0 \pmb Z_0^\intercal \big]
                        +  E\bigg[\sum_{j=1}^k \int_0^t \pmb Z_{s-} dA_s^j
                        \pmb Z_{s-}^\intercal  \nabla F_j(\pmb X_s)^\intercal                         
                         \bigg]
\\&    + E\bigg[\sum_{j= 1}^k 
\int_0^t  \nabla F_j(\pmb X_s) \pmb Z_{s-}d A^j _s \pmb Z_{s-}^\intercal           \bigg]
\\ & + E\bigg[ \int_0^t F(\pmb X_{s-}) d [ \pmb W] _s
                        F(\pmb X_{s-})^\intercal \bigg] \\
= &  E\big[\pmb Z_0 \pmb Z_0^\intercal \big]
                        +  \sum_{j=1}^k \int_0^tE\big[ \pmb Z_{s-}  \pmb Z_{s-}^\intercal
                        \big]                         \nabla F_j(\pmb X_s)^\intercal   
                        dA_s^j
\\&    + \sum_{j= 1}^k 
\int_0^t  \nabla F_j(\pmb X_s) E\big[\pmb Z_{s-} \pmb Z_{s-}^\intercal \big]      dA^j  
\\ & + \int_0^t F(\pmb X_{s-}) d [ \pmb W ] _s
                        F(\pmb X_{s-})^\intercal.  
                \end{align*}
                
We now show that the solutions of the SDE
\eqref{eq:XVarEst} converge in law to the covariance of $\pmb Z$. We first note
that
\cite[Theorem VI.6.26]{JacodShiryaev} implies that 
$\{[\pmb W^{(n)}] \}_n$ converges in law to  $[\pmb W]$ 
since $\{\pmb W^{(n)} \}_n$ is predictably uniformly tight. 
Moreover, since $\pmb A$ is continuous, \cite[Corollary VI 3.33]{JacodShiryaev} implies that $\{( \pmb {\hat A}^{(n)}, [\pmb W^{(n)}])  \}_n$ 
converges in law to $( \pmb {A}, [\pmb W])$. If this sequence was predictably uniformly tight, then we could apply 
\cite[Theorem IX.6.9]{JacodShiryaev} one more time to ensure that the solutions 
of the SDEs \eqref{eq:XVarEst} converge in law to $\pmb V$. 

Finally, note that  $\{( \pmb {\hat A}^{(n)}, [\pmb W^{(n)}] ) \}_n$ is P-UT
since both $\{\pmb {\hat A}^{(n)}\}_n$ and $\{[\pmb W^{(n)}]\}_n$ are P-UT.

\end{proof}

\subsection{Proof of Proposition \ref{prop:AalenPUT}}

\begin{proof}
        It is well known that $\pmb W^{(n)} := \sqrt n ( \pmb  A^{(n)}  - \pmb A ) $ %\int_0^\cdot \pmb \alpha_s ds)$ 
        converges weakly to a Gaussian process with independent increments, see \cite{Andersen}. To see that $\{\pmb W^{(n)} \}_n$ is P-UT, let 
        \begin{equation}
                \pmb \Gamma^{(n)}_t :=   \frac{\pmb U^{(n)\intercal}_{t-} \pmb Y^{(n)}_t \pmb U_{t-}^{(n)}  }{ n}, 
        \end{equation}
        and note that 
        \begin{align*}
                \Tr \langle \pmb W^{(n)} \rangle_{\mathcal T} = & \int_0^{\mathcal T}  \Tr \big( 
                \pmb  \Gamma^{(n)-1}_t 
\frac {\pmb U^{(n)\intercal}_{t-}\pmb  \lambda_t^{(n)}
                                \pmb Y^{(n)}_t
                               \pmb  U_{t-}^{(n)}  }{ n}
                         \pmb \Gamma^{(n)-1}_t 
                \big)
                dt.
        \end{align*}
From the submultiplicativity of the trace norm we have
\begin{align*}
        &  \Tr \big( 
                 \pmb \Gamma^{(n)-1}_t 
\frac {\pmb U^{(n)\intercal}_{t-} \pmb \lambda_t^{(n)}
                                \pmb Y^{(n)}_t
                                \pmb U_{t-}^{(n)}  }{ n}
                         \pmb \Gamma^{(n)-1}_t 
                \big)
                \\  \leq & 
                \Tr \big( 
                 \pmb \Gamma^{(n)-1}_t \big) \Tr \bigg(
\frac {\pmb U^{(n)\intercal}_{t-} \pmb \lambda_t^{(n)}
                                \pmb Y^{(n)}_t
                                \pmb U_{t-}^{(n)}  }{ n}
                       \bigg)\Tr \big(  \pmb \Gamma^{(n)-1}_t 
                \big). 
\end{align*}
On the other hand, we have that  \begin{equation}
      \Tr \bigg(
\frac {\pmb U^{(n)\intercal}_{t-} \pmb \lambda_t^{(n)}
                                \pmb Y^{(n)}_t
                                \pmb U_{t-}^{(n)}} { n}
                       \bigg) = \frac 1 n \sum_{k,i} (U_{t-}^{k,i})^2 Y_t^i
                       \lambda_t^i,                        
\end{equation}
and $E_P [\sup_{t \leq \mathcal{T}} | (U_{t-}^{k,i})^2 Y_t^i \lambda^i_t
   | ] < \infty$, so 
 Markovs inequality implies that 
\begin{equation}
        \lim_J \inf_n P \bigg(  \Tr \Big(
\frac {\pmb U^{(n)\intercal}_{t-} \pmb \lambda_t^{(n)}
                                \pmb Y^{(n)}_t
                                \pmb U_{t-}^{(n)}  }{ n} \Big)
                       \geq J  \bigg) = 0. 
\end{equation}
%        Therefore $  \Tr \big( \pmb \Gamma^{(n)-1}_t \frac {\pmb U^{(n)\intercal}_{t-} \pmb \lambda_t^{(n)}\pmb Y^{(n)}_t\pmb U_{t-}^{(n)}  }{ n}\pmb \Gamma^{(n)-1}_t \big)$.
  %$ satisfies $\ref{enum:PUTMartingale})$ of Lemma \ref{lemma:P-UTexplained}.
Finally, \cite[Proposition  II.5.3]{Andersen}  implies that \eqref{eq:PUTMart} of Lemma \ref{lemma:P-UTexplained} is satisfied.  

Note that $\{\pmb A^{(n)}\}_n$ is P-UT since $\pmb A^{(n)} = \frac {1}{\sqrt n} \pmb W^{(n)} +
        \pmb A_t$. To see that $\{[\pmb W^{(n)}]\}_n$ is P-UT, note that we have already
        seen that 
        the compensator of $[\pmb W^{(n)}]$, $\langle \pmb W^{(n)}\rangle$, satisfies 
        \eqref{eq:PUTMart} of Lemma \ref{lemma:P-UTexplained}. Moreover, note
        that 
\begin{align*}
        \Tr \big(\langle  [\pmb W^{(n)}] - \langle \pmb W^{(n)}\rangle \rangle_{\mathcal{T}}\big) = 
         \Tr \big(\langle \pmb W^{(n)} \rangle_{\mathcal{T}}\big), 
\end{align*}
        so $\eqref{eq:PUTMart}$ of Lemma 
\ref{lemma:P-UTexplained} is satisfied for $\{   [\pmb W^{(n)}] - \langle
\pmb W^{(n)}\rangle \}_n$ as well, which therefore implies that  
$\{[\pmb W^{(n)}]\}_n$ is P-UT.

\end{proof}

%\section*{Simulation details}
%For all the simulations displayed in Figure \ref{fig:panelPlotParameters1} - \ref{fig:coverageCrossingHazards} we considered a study period $[0,40]$, and two sets of hazards: constant, and linear. The constant hazards we considered were
%\begin{align*}
%  alpha_t^0 &= 0.06\\
%  alpha_t^1 &= 0.04,
%\end{align*}
%while the linear (crossing) hazards were
%\begin{align*}
%	\alpha_t^0 &=0.14 - 0.003t \\
%    \alpha_t^1 &=0.02 + 0.003t
%\end{align*}
%
%In all but Figure \ref{fig:coverageCrossingHazards} we used constant hazards. Using the notation in the examples of Section \ref{section:parameters}, we assigned hazards as follows:
%\begin{itemize}
%    \item $\alpha^0$: Relative survival group 1, Cumulative sensitivity group 0, LED and LER group 1, Cumulative incidence event hazard.
%    \item $\alpha^1$: Survival, Relative survival group 0, Cumulative sensitivity group 1, LED and LER group 2.
%\end{itemize}

\section*{Appendix: Other plugin variance expressions}

\subsection*{Variance for LER}
Here, the plugin variance reads
\begin{align*}
    \pmb{\hat V}_t &= \pmb{\hat V}_{\tau_{k-1}} + \Big( \pmb{\hat V}_{\tau_{k-1}} G_{\tau_{k-1}}^\top + G_{\tau_{k-1}}\pmb{\hat V}_{\tau_{k-1}} \Big) \Delta \tau_{k} \\
    &- \Big( \pmb{\hat V}_{\tau_{k-1}} e_{2,2}%\begin{pmatrix} 0&0&0&0&0\\0&-1&0&0&0\\0&0&0&0&0\\0&0&0&0&0\\0&0&0&0&0
%    \end{pmatrix} + \begin{pmatrix} 0&0&0&0&0\\0&-1&0&0&0\\0&0&0&0&0\\0&0&0&0&0\\0&0&0&0&0
    %\end{pmatrix}
    +e_{2,2}\pmb{\hat V}_{\tau_{k-1}} \Big) \Delta \hat{A}_{\tau_k}^1 \\
    &- \Big( \pmb{\hat V}_{\tau_{k-1}} e_{3,3}%\begin{pmatrix} 0&0&0&0&0\\0&0&0&0&0\\0&0&-1&0&0\\0&0&0&0&0\\0&0&0&0&0
%    \end{pmatrix} + \begin{pmatrix} 0&0&0&0&0\\0&0&0&0&0\\0&0&-1&0&0\\0&0&0&0&0\\0&0&0&0&0
    %\end{pmatrix}
    +e_{3,3}\pmb{\hat V}_{\tau_{k-1}} \Big) \Delta \hat{A}_{\tau_k}^2 \\
    & + nF_{\tau_{k-1}} \Delta\begin{pmatrix} 0\\\hat{A}_{\tau_k}^1\\\hat{A}_{\tau_k}^2 \end{pmatrix}\Delta\begin{pmatrix} 0 & \hat{A}_{\tau_k}^1 & \hat{A}_{\tau_k}^2 \end{pmatrix} F_{\tau_{k-1}}^\top,
\end{align*}
where $G$ and $F$ is given by
\begin{align*}
    G_{\tau_{k-1}} &= \begin{pmatrix} 0&\frac{1}{\hat{R}_{\tau_{k-1}}^2}&-\frac{\hat{R}_{\tau_{k-1}}^1}{(\hat{R}_{\tau_{k-1}}^2)^2}&-\frac{\hat{S}_{\tau_{k-1}}^2}{(\hat{R}_{\tau_{k-1}}^2)^2}&\frac{2 \hat{S}_{\tau_{k-1}}^2 \hat{R}_{\tau_{k-1}}^1 - \hat{S}_{\tau_{k-1}}^1\hat{R}_{\tau_{k-1}}^2 }{(\hat{R}_{\tau_{k-1}}^2)^3}\\0&0&0&0&0\\0&0&0&0&0\\0&1&0&0&0\\0&0&1&0&0\end{pmatrix} \\
    F_{\tau_{k-1}} &= \begin{pmatrix}
        \frac{\hat{S}_{\tau_{k-1}}^1 \hat R_{\tau_{k-1}}^2 - \hat{S}_{\tau_{k-1}}^2 \hat R_{\tau_{k-1}}^1}{(\hat R_{\tau_{k-1}}^2)^2} & 0 & 0  \\
        0 & -\hat{S}_{\tau_{k-1}}^1 & 0  \\
        0 & 0 & -\hat{S}_{\tau_{k-1}}^2  \\
        \hat{S}_{\tau_{k-1}}^1 & 0 & 0 \\
        \hat{S}_{\tau_{k-1}}^2 & 0 & 0
    \end{pmatrix}
\end{align*}

\subsection*{Variance for cumulative sensitivity and specificity}
In this case the plugin variance is
\begin{align*}
    \pmb{\hat V}_t &= \pmb{\hat V}_{\tau_{k-1}} + \Big( \pmb{\hat V}_{\tau_{k-1}} G_{\tau_{k-1}}^{1\top} + G_{\tau_{k-1}}^1 \pmb{\hat V}_{\tau_{k-1}} \Big) \Delta \hat{A}_{\tau_{k}}^1\\
    &+ \Big( \pmb{\hat V}_{\tau_{k-1}} G_{\tau_{k-1}}^{2\top} + G_{\tau_{k-1}}^{2} \pmb{\hat V}_{\tau_{k-1}} \Big)\hat{A}_{\tau_{k}}^0 \\
    &+n F_{\tau_{k-1}} \Delta\begin{pmatrix} \hat{A}_{\tau_k}^1\\\hat{A}_{\tau_k}^0 \end{pmatrix}\Delta\begin{pmatrix} \hat{A}_{\tau_k}^1 & \hat{A}_{\tau_k}^0 \end{pmatrix} F_{\tau_{k-1}}^\top.
\end{align*}
Here we have defined
    \begingroup\makeatletter\def\f@size{8}\check@mathfonts\begin{align*}
        G^1_{\tau_{k-1}} &= \begin{pmatrix} -1&0&0&0\\0&0&0&0\\
            \frac{\hat{W}_{\tau_{k-1}}^2 (1-\hat{V} _{\tau_{k-1}})(\hat U_{\tau_{k-1}} - 2 ) }{ \gamma \hat{U}^3 _{\tau_{k-1}} } & \frac{ \hat{W}^2 _{\tau_{k-1}} (\hat{U} _{\tau_{k-1}} - 1) }{\gamma \hat{U}^2 _{\tau_{k-1}} } & \frac{2 \hat{W} _{\tau_{k-1}} (1 - \hat{U} _{\tau_{k-1}})(1- \hat{V} _{\tau_{k-1}})}{\gamma \hat{U}^2 _{\tau_{k-1}}} & 0 \\
            -\frac{\gamma \hat{X}^2 _{\tau_{k-1}}}{\hat{V} _{\tau_{k-1}}} & -\frac{\gamma \hat{X}^2 _{\tau_{k-1}} (1-\hat{U}_{\tau_{k-1}})}{\hat{V}^2 _{\tau_{k-1}}} & 0 & \frac{2 \gamma \hat{X} _{\tau_{k-1}} ( 1 - \hat{U} _{\tau_{k-1}}) }{\hat{V} _{\tau_{k-1}}}
        \end{pmatrix}, \\
        G_{\tau_{k-1}}^2 &= \begin{pmatrix} 
            0&0&0&0\\0&-1&0&0\\
            \frac{\hat W_{\tau_{k-1}}^2  \hat V _{\tau_{k-1}}}{\gamma \hat U^2 _{\tau_{k-1}}} & -\frac{\hat W_{\tau_{k-1}} }{\gamma \hat U_{\tau_{k-1}} } & -\frac{2 \hat W_{\tau_{k-1}} \hat V_{\tau_{k-1}}}{ \gamma \hat U_{\tau_{k-1}} } & 0 \\            \frac{\gamma \hat{X}^2 _{\tau_{k-1}}}{\hat{V} _{\tau_{k-1}}} & \frac{\gamma \hat{X}^2 _{\tau_{k-1}} (1- \hat{U} _{\tau_{k-1}})}{\hat{V}^2 _{\tau_{k-1}}} & 0 & - \frac{2 \gamma \hat{X} _{\tau_{k-1}} ( 1 - \hat{U} _{\tau_{k-1}}) }{\hat{V} _{\tau_{k-1}}}
        \end{pmatrix},
        \\
        F_{\tau_{k-1}} &= \begin{pmatrix}
               1 -\hat U_{\tau_{k-1}}  & 0  \\
               0  & -\hat V_{\tau_{k-1}} \\
               \frac{\hat W^2_{\tau_{k-1}} ( 1- \hat V_{\tau_{k-1}})( 1-
                         \hat U_{\tau_{k-1}}) }{\gamma \hat U_{\tau_{k-1}}^2}  
               &
                - \frac{\hat W^2_{\tau_{k-1}} \hat V_{\tau_{k-1}}\hat U_{\tau_{k-1}} } { \gamma \hat U_{\tau_{k-1}}^2 } 
                \\ 
                \frac{ \gamma \hat X_{\tau_{k-1}}^2 (1-
            \hat    U_{\tau_{k-1}}) }{\hat V_{\tau_{k-1}} }  & -
                \frac{\gamma \hat X_{\tau_{k-1}}^2 (1-
                \hat U_{\tau_{k-1}}) }{\hat V_{\tau_{k-1}} }
        \end{pmatrix}.
    \end{align*}\endgroup

%\begin{figure}[ht]
%    \centering
%  \includegraphics[width=\textwidth]{panelPlotParameters1}
%    \caption{Plugin estimates with population sizes $n=500,1000,$ and $2000,$ along with 95 \% pointwise confidence intervals. Exact parameters is shown in red, dotted lines.}
%    \label{fig:panelPlotParameters1}
%\end{figure}

%begin{figure}[ht]
%   \centering
% \includegraphics[width=\textwidth]{panelPlotParameters2}
%   \caption{Plugin estimates with population sizes $n=500,1000,$ and $2000,$ along with 95 \% pointwise confidence intervals. Exact parameters is shown in red, dotted lines.}
%   \label{fig:panelPlotParameters2}
%end{figure}

%begin{figure}[ht]
%   \centering
   %\includegraphics[width=\textwidth]{jointConvConst}
%   \caption{Convergence of selected plugin estimators; parameter to the left and variances to the right. From left to right we included order $1$ and  $2.3$ for comparison.}
%   \label{fig:jointConvConst}
%end{figure}

%begin{figure}[ht]
%   \centering
   %\includegraphics[width=\textwidth]{constHazardCoverage.pdf}
%   \caption{Estimated mean coverage for selected parameters with constant hazards. The dotted red line indicates the confidence level. }
   %\label{fig:coverageConstantHazards}
%end{figure}

%begin{figure}[ht]
%   \centering
   %\includegraphics[width=\textwidth]{crossingHazardCoverage.pdf}
%   \caption{Estimated mean coverage for selected parameters with linear (crossing) hazards. The dotted red line indicates the confidence level. %
   %\label{fig:coverageCrossingHazards}
%end{figure}

\bibliography{references}
\bibliographystyle{ieeetr}
\end{document}